\documentclass[amsmath,amssymb,prb,letterpaper, twocolumn, floatfix, superscriptaddress, longbibliography]{revtex4-2}
\usepackage{bm,graphicx,hyperref}
\hypersetup{
  breaklinks = {true},
  citecolor = {blue},
  colorlinks = {true},
  linkcolor = {red},
}

\begin{document}

\title{Non-equilibrium angular momentum selectivity and filtering in chiral carbon nanotubes}

\author{Sergio Shmayev}
\affiliation {School of Engineering, Mackenzie Presbyterian University, S\~ao Paulo - 01302-907, Brazil}
\affiliation {MackGraphe Graphene and Nanomaterials Research Institute, Mackenzie Presbyterian University, S\~ao Paulo -01302-907, Brazil}

\author{D. R. da Costa}
\affiliation{Departamento de Física, Universidade Federal do Ceará, 60455-900, Fortaleza, CE, Brazil}

\author{D. A. Bahamon}
\email{dario.bahamon@mackenzie.br}
\affiliation {School of Engineering, Mackenzie Presbyterian University, S\~ao Paulo - 01302-907, Brazil}
\affiliation {MackGraphe Graphene and Nanomaterials Research Institute, Mackenzie Presbyterian University, S\~ao Paulo -01302-907, Brazil}

\begin{abstract}
Carbon nanotubes (CNTs) constitute a highly tunable platform for probing the interplay between structural chirality and quantum transport in quasi-one-dimensional systems. Here, we perform a systematic study of the non-equilibrium orbital response across a broad set of metallic and semiconducting chiral CNTs. We find that the orbital Edelstein susceptibility depends strongly on both chirality and nanotube diameter, revealing that the orbital response cannot be captured by a universal scaling law. Instead, distinct families of CNTs emerge, forming characteristic orbital-response branches uniquely determined by the chiral wrapping vector. We further investigate the role of metallic contacts on orbital-current generation and orbital selectivity. While metallic CNTs rapidly recover their intrinsic orbital response away from the contact region, semiconducting CNTs display pronounced oscillatory behavior arising from interference between transport channels carrying different angular momenta injected by wide-band metallic contacts. Finally, by incorporating angular correlations into the contact self-energy, we demonstrate that chiral CNTs can operate as efficient orbital-angular-momentum filters, selectively transmitting orbitally textured electronic states in accordance with the crystal angular momentum of the propagating bands. 
\end{abstract}

\maketitle

The inherent lack of mirror and spatial-inversion symmetries in chiral systems gives rise to a variety of quantum transport phenomena that intimately link the structural handedness of a material to its electronic and magnetic responses~\cite{Binghai_review, Chirality_quantumLeap}. Among the most extensively studied manifestations, yet still not fully understood, are the chiral-induced spin selectivity (CISS) effect and electrical magnetochiral anisotropy (eMChA)~\cite{Review_CISS, NRTokura, doi:10.1021/acs.jpclett.3c02546}. In the CISS effect, electrons transmitted through a chiral material acquire a net spin polarization whose sign is strictly dictated by the enantiomeric form of the structure. This observation has sparked intense debate about its microscopic origin, given that the measured polarization magnitude far exceeds naive estimates based on spin-orbit coupling~\cite{Naaman2019, Evers2022, Liu2021spinfilter}. In the latter, eMChA manifests as a nonreciprocal correction to the electrical resistance that depends simultaneously on the relative orientation of the charge current and the applied external magnetic field, thereby breaking Onsager reciprocity and providing a direct macroscopic fingerprint of the underlying structural chirality~\cite{Rikken2001, Pop2014, Yokouchi2017, Ideue2017}. More recently, the concept of chirality as an active transport observable has been extended beyond the spin channel: chirality-driven orbital polarization, nonlocal voltage probes of handedness in twisted bilayers, and other novel transport signatures have emerged as active areas of research, broadening the scope of chirality-sensitive phenomena \cite{BahamonChV3, MatheusCount, InuiCrNb3S6}.

Regarding the orbital degree of freedom and its manipulation, an emerging field known as orbitronics \cite{Rev_orbitronics, Go2021orbitaltorque}, recent studies have demonstrated that when an electric current flows through a chiral material, an orbital magnetic moment is generated whose sign and magnitude depend on the structural handedness of the system~\cite{doi:10.1021/acs.nanolett.7b04300, Binghai_DNA, doi:10.1021/acsnano.3c03893, https://doi.org/10.1002/adma.202418040}. This chirality-induced orbital magnetization constitutes an orbital analog of the Edelstein effect, whereby a polar vector (the applied electric field) is converted into an axial vector (the orbital moment) through the mediation of the chiral crystal symmetry~\cite{doi:10.1021/acs.nanolett.7b04300, Bhowal2022}. By analogy with the spin polarization observed in the CISS effect, this generation of orbitally textured non-equilibrium states has been termed chirality-induced orbital selectivity (CIOS)~\cite{Gobel_CIOS}. Notably, CIOS provides a compelling route to large, gate-tunable orbital magnetizations without relying on relativistic spin-orbit coupling, making it particularly attractive for light-element materials~\cite{Go2021orbitaltorque, Salemi2022}. However, the vast majority of previous theoretical studies have relied on idealized low-dimensional model Hamiltonians~\cite{Gobel_CIOS}. In realistic materials, microscopic studies of CIOS remain relatively limited, and only recently has the orbital Edelstein susceptibility been investigated in metallic chiral CNTs \cite{Gobel_CNT}.

While the generation of orbital magnetic moments discussed above is closely related to current-driven orbital transport, the appearance of orbital magnetism in CNTs is by no means a new phenomenon. Large orbital magnetic moments, reaching several Bohr magnetons per electron,  have been experimentally observed in CNTs subjected to an axial magnetic field via single-nanotube magnetoconductance and scanning-probe  spectroscopy~\cite{RevModPhys.79.677, RevModPhys.87.703, Minot2004, Jarillo2005}, and the formation of azimuthal currents, which endow the nanotube with a tube-axis magnetic moment analogous to a nanoscale quantum solenoid, exhibiting a rich dependence on the chiral wrapping angle, has long been recognized \cite{PhysRevLett.76.2121, PhysRevB.60.13885, PhysRevB.78.233405}. However, these descriptions are generally based either on semiclassical models that treat the azimuthal motion as a free-electron ring or on the electronic band structure of graphene formulated within the zone-folding approximation. More recently, attention has shifted toward the interplay between the crystal angular momentum as a conserved quantum number imposed by the discrete rotational symmetry $C_N$ of the nanotube, the valley degree of freedom, and the finite nanotube length \cite{PhysRevB.92.075433, PhysRevB.91.235442, PhysRevB.93.195442, Marganska2019}. These analyses demonstrate that, while CNTs may appear to share universal transport features at a coarse energy scale set by the subband spacing, each specific $(n,m)$ chirality possesses a unique and experimentally distinguishable transport fingerprint encoded in the angular-momentum composition of its propagating channels~\cite{RevModPhys.87.703}.

Motivated by the need to bridge the gap between idealized model predictions and the transport physics of realistic chiral nanostructures, in the first part of this work, we systematically investigate the non-equilibrium current-driven orbital selectivity of both metallic and semiconducting CNTs. To systematically assess the combined influence of nanotube geometry (diameter) and chirality (chiral angle), we analyze a total of 86 distinct CNTs, including 25 metallic and 61 semiconducting systems \cite{tubegen2011}. Since practical CNT devices necessarily require electrical contacts, we also examine the impact of metallic contacts on the orbital selectivity and on the efficiency of orbital-current generation. In the second part of the work, we explore the orbital-angular-momentum filtering capabilities of CNTs by introducing a wide-band contact model that includes angular correlations among atoms coupled to the metallic electrode, thereby enabling the injection of electrons with a controlled orbital texture. Our central findings are threefold and reveal that chirality plays a central role in determining the non-equilibrium orbital response. First, the orbital Edelstein susceptibility cannot be described solely in terms of the nanotube diameter, as the data do not collapse onto a universal curve; rather, it segregates into distinct branches uniquely determined by the chiral indices $(n,m)$. Second, the spatial evolution of the orbital magnetization near the contacts reveals a fundamental asymmetry between metallic and semiconducting CNTs: metallic CNTs rapidly recover their bulk orbital response within a few unit cells of the lead, whereas semiconducting CNTs develop long-range oscillations arising from quantum beating between modes of different crystal angular momentum. Third, the filtering efficiency approaches unity for appropriately chosen contact correlations, establishing chiral CNTs as efficient orbital-angular-momentum filters, relevant for effective components for angular-momentum-resolved orbitronic architectures, selectively transmitting electronic states according to their angular-momentum character.  While the orbital Edelstein effect in metallic CNTs was recently introduced in Ref.~[\onlinecite{Gobel_CNT}], the present work extends that analysis in several directions: we encompass both metallic and semiconducting families across 86 chiralities, uncover the family-dependent branch structure of the susceptibility, resolve the spatial buildup of orbital magnetization near realistic contacts, and demonstrate angular-momentum filtering via orbitally textured injection.

\begin{center}
\begin{figure}[t]
\scalebox{0.85}{\includegraphics{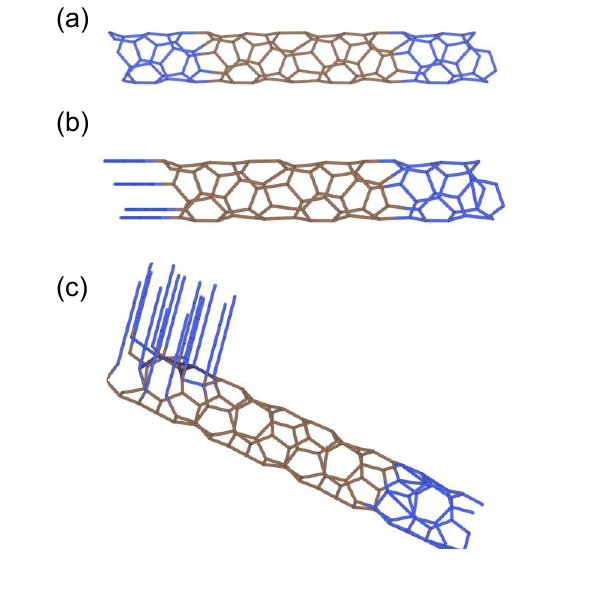}}
\caption{\textbf{Schematic representation of the CNT transport devices illustrating the three different contact geometries considered in this work.} (a)~Ideal semi-infinite CNT leads: both left and right electrodes are pristine extensions of the same nanotube, preserving the full helical symmetry and injecting all angular-momentum channels uniformly. (b) End-contact geometry: the left electrode is replaced by a wide-band metallic lead coupled to all atoms at the leftmost cross-section of the nanotube via the self-energy $\Sigma_L = -i|t|$. (c) Side-contact geometry: the metallic lead couples only to the atoms satisfying $y \geq 0$ within the first unit cell, thereby breaking the discrete rotational symmetry $C_N$ of the contact interface and selectively addressing a subset of angular-momentum channels. In all panels, the leads are highlighted in blue.}
\label{fig:fig1} 
\end{figure}
\end{center}

\section{Model}
\label{sec:model}

A single-wall CNT can be conceptually constructed by and viewed as a section of a graphene sheet that is cut and subsequently rolled into a cylindrical shape \cite{bookSaito, DresselhausBook} [see Fig.~\ref{fig:fig1}(a)]. The orientation and size of the selected graphene section are specified by the integer pair $(n,m)$, which determines the mutually perpendicular chiral vector $\mathbf{C}_h = n\mathbf{a}_1 + m\mathbf{a}_2$ and the translational vector $\mathbf{T}$. The chiral vector determines the nanotube circumference $|\mathbf{C}_h| = a\sqrt{n^2 + nm + m^2}$ and, consequently, the tube diameter $d_t = |\mathbf{C}_h|/\pi$, and is constructed from the graphene lattice vectors, $\mathbf{a}_1 = \left( \frac{3a}{2}, -\frac{\sqrt{3}a}{2} \right)$, $\mathbf{a}_2 = \left( \frac{3a}{2}, \frac{\sqrt{3}a}{2} \right)$, where $a = 1.42\,\text{\AA}$ is the carbon--carbon bond length. The translation vector, which is aligned with the nanotube axis, is given by $\mathbf{T} = \frac{2n+m}{d_R}\mathbf{a}_1 - \frac{2m+n}{d_R}\mathbf{a}_2$, where $d_R$ is the greatest common divisor of \(2m+n\) and \(2n+m\). Owing to the hexagonal point-group symmetry of graphene, all physically distinct chiralities are contained within the sector $0 \leq m \leq n$. Under this convention, the orientation of the hexagonal lattice relative to the CNT axis is characterized by the chiral angle $\theta$, which satisfies $ 0 \leq |\theta| \leq 30^\circ$. The limiting chiral angle $\theta = \arctan\!\left[\sqrt{3}\,m/(2n+m)\right]$ range is from $\theta = 0^\circ$ (zigzag) to $\theta = 30^\circ$ (armchair), with all intermediate values corresponding to genuinely chiral nanotubes that lack any mirror plane containing the tube axis~\cite{bookSaito, WhiteMintmire1998}. It is precisely this absence of mirror symmetry that enables the chirality-dependent orbital phenomena investigated in this work.

Once the nanotube geometry is specified, we describe its electronic structure within a nearest-neighbor tight-binding framework, retaining a single $p_z$ orbital per carbon atom and a hopping integral $|t| = 3$~eV~\cite{bookSaito, Reich2004book}. This minimal model faithfully reproduces the low-energy band structure of CNTs, including the metallic or semiconducting character dictated by the condition $(n - m) \bmod 3 = 0$~\cite{RevModPhys.79.677}, and captures the essential subband structure and angular-momentum quantum numbers required for the orbital transport analysis presented below. We further note that a uniform, symmetry-preserving next-nearest-neighbor hopping, which connects atoms within the same sublattice, preserves the discrete rotational symmetry $C_N$ of the nanotube. Consequently, the crystal angular momentum $\mathfrak{m}$ remains a good quantum number, and the angular-momentum selection rules derived in Section~\ref{sec:filtering} imposed by $C_N$ are unaffected. The main quantitative effect of second-neighbor hopping is to break electron-hole symmetry, including a shift of the charge-neutrality energy and introducing an asymmetry between the electron- and hole-doped orbital responses without altering the fundamental chirality-dependent branch structure.

\section{Chirality-induced selectivity of angular momentum}

In the first part of this work, we evaluate the non-equilibrium orbital magnetic moment induced in a chiral CNT by an applied longitudinal electric field. Within the Landauer--B\"uttiker formalism combined with the non-equilibrium Green's function (NEGF) technique~\cite{Datta1995, bahamon2020emergent}, the bond electric current flowing from site $i$ to site $j$ under a finite source--drain bias $V_{SD}$ is given by
\begin{equation}
I_{i\rightarrow j}
=
\frac{2e}{h}
\int_{E_F - V_{SD}/2}^{E_F + V_{SD}/2}
dE
\,
\mathrm{Im}
\left[
t_{ij}G_{ji}^{<}(E)
\right].
\label{eq:Iij}
\end{equation}
Where  $t_{ij} = t$ is the nearest-neighbor hopping integral (vanishing otherwise), and $G_{ji}^{<}(E)$ denotes the lesser Green's function encoding the non-equilibrium occupation of electronic states between sites $j$ and $i$~\cite{Datta1995, HaugJauho2008}. Throughout this work, the gate voltage $V_g$ rigidly shifts the Fermi level at which the transport integrals are evaluated, \textit{i.e.}, $E_F \equiv V_g$ in Eq.~\eqref{eq:Iij}, while the Hamiltonian remains unchanged. Here, we operate in the linear-response regime by setting $V_{SD} = 100~\mu\text{eV}$, which is small compared to all relevant energy scales (subband spacing, hopping integral) yet sufficiently large to ensure numerical convergence of the current distribution. We emphasize that the NEGF formalism is employed here not because of strong non-equilibrium effects---the small bias $V_{SD} = 100~\mu$eV places the system firmly in the linear-response regime---but rather because it provides the natural framework for computing spatially resolved bond currents in finite-size devices with broken translational symmetry~\cite{Datta1995, HaugJauho2008}. In particular, the lesser Green's function $G^<_{ji}(E)$ gives direct access to the local current distribution [Eq.~\eqref{eq:Iij}] and hence to the unit-cell-resolved orbital magnetic moment profiles presented in Section~\ref{sec:metalcontact}, which cannot be obtained from a transmission-only calculation. We note that the real-space lesser Green's function $G^<_{ji}(E)$ implicitly contains the full mode structure of the nanotube: it can be formally decomposed as $G^<_{ji}(E) = i\sum_\nu \psi_\nu(j)\,\psi^*_\nu(i)\,f_\nu(E)$, where $\nu$ labels all propagating Bloch states---including band index and longitudinal wave vector $k_z$---at energy $E$, and $f_\nu(E)$ encodes the non-equilibrium occupation set by the contact self-nergies~\cite{Datta1995, HaugJauho2008}. The real-space formulation is therefore fully equivalent to the reciprocal-space expressions employed in Refs.~[\onlinecite{doi:10.1021/acs.nanolett.7b04300, Gobel_CNT}], while being naturally suited to finite-size devices where translational symmetry is broken by the contacts.

The local orbital magnetic moment induced by the electronic current distribution is obtained from the discrete lattice analog of the continuum expression $\mathbf{m} = \frac{1}{2}\int \mathbf{r}\times\mathbf{J}\,dV$, yielding~\cite{bahamon2020emergent, Peierls1933}
\begin{equation}
\mathbf{m}
=
\frac{1}{2}
\sum_{\langle i,j \rangle}
I_{i\rightarrow j}
\,
\left(
\mathbf{r}_i \times \mathbf{r}_j
\right),
\label{eq:orbital_moment}
\end{equation}
where $\mathbf{r}_i$ and $\mathbf{r}_j$ are the position vectors of the carbon sites and the summation runs over all nearest-neighbor pairs $\langle i,j\rangle$. This expression captures the orbital magnetization arising from the circulation of bond currents around the nanotube circumference, \textit{i.e.} the microscopic origin of the transverse orbital magnetic moment in chiral CNTs. Importantly, for achiral (armchair or zigzag) CNT, the azimuthal components of the bond currents cancel by symmetry, and $\mathbf{m}$ vanishes identically in the absence of an external magnetic field, confirming that a nonzero transverse orbital moment is a direct fingerprint of structural chirality.

To characterize the non-equilibrium orbital magnetic response along the nanotube axis ($z$-direction), we define the orbital Edelstein susceptibility as \cite{doi:10.1021/acs.nanolett.7b04300, Gobel_CNT}
\begin{equation}
\chi_z^{L_z} = \frac{m_z}{\mathcal{E}_z},
\label{eq:edelstein}
\end{equation}
where $m_z$ is the axial component of the orbital magnetic moment obtained from Eq.~\eqref{eq:orbital_moment}, and the longitudinal electric field is 
\begin{equation}
\mathcal{E}_z = \frac{V_{SD}}{L},
\end{equation}
with $L$ denoting the CNT length. This definition follows the convention introduced by Yoda \textit{et al.}~\cite{doi:10.1021/acs.nanolett.7b04300} for the orbital Edelstein effect, wherein a polar vector (the applied field) generates an axial pseudo-vector (the orbital moment) through the mediation of the chiral crystal structure. The susceptibility $\chi_z^{L_z}$ thus encapsulates the efficiency with which the structural chirality converts a longitudinal charge current into a transverse orbital magnetization.

The formalism described above is general and applies independently of the contact model employed. To assess the role of realistic metallic leads, we replace the pristine semi-infinite CNT electrode on the left side by a wide-band metallic contact, implemented through a retarded self-energy $\Sigma_L = -i\,|t|$~\cite{Bahamon, Datta1995} assigned to the subset of carbon atoms of the CNT directly coupled to the lead. This choice corresponds to a featureless metallic reservoir with a constant density of states, ensuring that the injected current carries no preferential angular-momentum texture, a deliberate design that isolates the CNT's intrinsic filtering properties from any contact-induced bias. Two relevant contact geometries are considered, as shown in Figs.~\ref{fig:fig1}(b) and \ref{fig:fig1}(c). For the \textit{end-contacted} geometry [Fig.~\ref{fig:fig1}(b)], the CNT is connected through its left edge, and therefore the onsite energies of the leftmost atoms are modified by $\Sigma_L$, mimicking a situation where the nanotube terminates at a planar metallic electrode~\cite{Nemec2006}. In the \textit{side-contacted} geometry [Fig.~\ref{fig:fig1}(c)], only atoms satisfying $y > 0$ are coupled to the lead, representing a scenario in which the nanotube rests on top of a metallic substrate and couples to it over a fraction of its circumference, consequently acquiring the $\Sigma_L$ contribution~\cite{Tersoff1999, Chen2005contact}. This distinction is physically significant because the side contact breaks the residual rotational symmetry of the end-contacted case, thereby modifying the set of angular-momentum channels that are efficiently injected into the tube.

For the systematic survey of chirality-dependent orbital responses, we consider all chiral indices within the sector $0 \leq m \leq n$ subject to the constraint $n + m \leq 20$, yielding a set of 86 distinct CNTs that spans a broad range of diameters and chiral angles for both metallic and semiconducting families~\cite{tubegen2011}. This upper bound on $n+m$ ensures computational tractability while still covering diameters up to $\sim 1.6$~nm, well within the range accessible to chemical vapor deposition synthesis~\cite{Zhang2017CVD}.  Since our primary objective is to isolate the role of the chiral wrapping vector in determining the orbital response, curvature-induced $\sigma$--$\pi$ rehybridization effects, which become appreciable only for very small diameters ($d_t \lesssim 0.5$~nm)~\cite{Blase1994, Reich2004book}, are not included in the present study. This approximation is well justified for the majority of our dataset ($d > 0.5$~nm), where curvature corrections are negligible compared to the primary subband spacing (see Appendix~\ref{sec.appendix}). We also emphasize that the $\sigma$ bands, formed by the in-plane $sp^2$ orbitals, lie more than 3~eV below (bonding) and above (antibonding) the charge-neutrality point~\cite{bookSaito, Reich2004book} and therefore do not participate in transport within the energy window considered here ($|V_g| \lesssim 1$~eV). 

\subsection{Metallic Carbon Nanotubes }

\begin{center}
\begin{figure}[t]
\scalebox{0.85}{\includegraphics{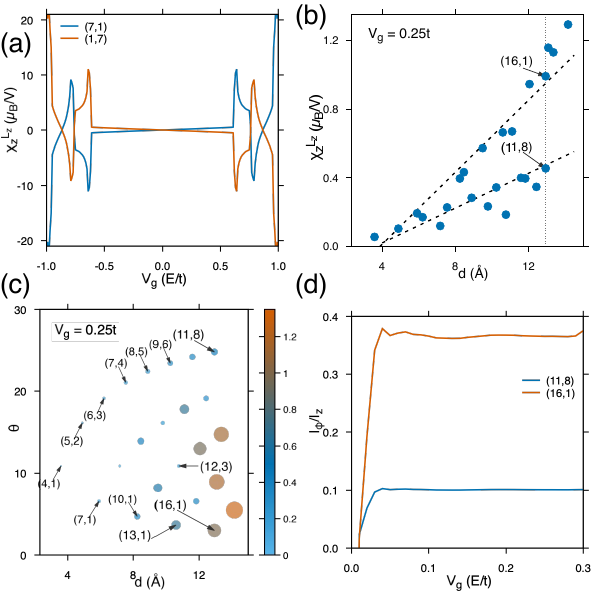}}
\caption{\textbf{Orbital Edelstein susceptibility of metallic chiral CNTs.} All values of $\chi_z^{L_z}$ are normalized by the respective unit-cell length $|\mathbf{T}|$ (see text for details). (a)~$\chi_z^{L_z}$ as a function of gate voltage for the enantiomeric pair $(7,1)$ and $(1,7)$, its mirror image, illustrating the sign reversal expected for a chirality-odd observable. (b)~Normalized orbital Edelstein susceptibility evaluated at $V_g = 0.25\,t$ as a function of nanotube diameter for all 25 metallic CNTs studied; the data do not collapse onto a single curve, revealing the absence of a universal diameter scaling. (c)~Bubble-chart representation of $\chi_z^{L_z}$ in the diameter--chiral-angle parameter space; the bubble area is proportional to $|\chi_z^{L_z}|$, highlighting the formation of distinct chirality-dependent branches. (d)~Ratio of the azimuthal to axial bond-current components, $I_\phi/I_z$, for the $(16,1)$ and $(11,8)$ CNTs---two chiralities sharing the same diameter ($d = 12.93~\text{\AA}$) but markedly different chiral angles---demonstrating that the orbital response is governed by the wrapping geometry rather than the tube radius alone.}
\label{fig:Fchimetal} 
\end{figure}
\end{center}

We begin by analyzing metallic CNTs, defined by the condition 
$(n - m)\,\mathrm{mod}\,3 = 0$, which ensures that the Dirac cones of the parent graphene sheet are sampled by the allowed transverse wave vectors~\cite{RevModPhys.79.677, bookSaito}. Figure~\ref{fig:Fchimetal}(a) displays the orbital Edelstein susceptibility $\chi_z^{L_z}$ as a function of the gate voltage $V_g$ for a representative metallic CNT. To enable a meaningful comparison across CNTs of different diameters and chiral angles, which possess widely varying numbers of atoms per unit cell and translational periods $|\mathbf{T}|$, the susceptibility is normalized per unit cell and expressed in units of $\mu_B/\mathrm{V}$, where $\mu_B$ denotes the Bohr magneton, as the convention reported in Ref.~[\onlinecite{Gobel_CNT}]. In the low-energy regime, where only the two linearly dispersing subbands at the $K$ and $K'$ valleys contribute to transport, $\chi_z^{L_z}$ exhibits a smooth dependence on the gate voltage, being approximately linear in $V_g$ around charge neutrality and evolving into a parabolic behavior ($\sim V_g^2$) as the Fermi level approaches the band edge of the first higher subband. This smooth behavior reflects the fact that, within a single pair of linear bands, the orbital moment is generated by the helical winding of the Bloch states around the nanotube circumference, whose magnitude scales monotonically with the longitudinal momentum~\cite{PhysRevLett.76.2121, PhysRevB.78.233405}. Note that as the Fermi level moves to higher energies, additional transport channels become available, thus, in turn, giving rise to oscillatory features in the susceptibility \cite{PhysRevB.78.233405, Gobel_CNT}. This multi-channel interference is a direct manifestation of the subband-resolved orbital texture inherent to chiral CNTs. As expected from the transformation properties of an axial vector under mirror reflection, nanotubes with opposite chiralities, such as $(7,1)$ and its enantiomer (mirror image) $(1,7)$, exhibit orbital Edelstein susceptibilities of equal magnitude but opposite signs, providing a stringent consistency check of the calculation~\cite{Gobel_CIOS, Gobel_CNT}. We note that the sign reversal $\chi_z^{L_z}(n,m) = -\chi_z^{L_z}(m,n)$ holds at every gate voltage, confirming that the orbital response is a true chirality-odd observable. A remark concerning the units is in order. The susceptibility $\chi_z^{L_z}$ reported per unit cell must be multiplied by the translational period $|\mathbf{T}|$ to yield the physical (length-intensive) Edelstein susceptibility. For the $(7,1)$ CNT at $V_g = 0.25\,t$, we obtain $\chi_z^{L_z} \approx 0.19~\mu_B/\mathrm{V}$ per unit cell. Given that $|\mathbf{T}| \approx 10.7~\text{\AA}$ for this chirality, the corresponding physical susceptibility is $\chi_z^{L_z} \approx 0.19 \times 10.7 \times 10^{-10} \approx 2 \times 10^{-10}~\mu_B\,\mathrm{m/V}$, a value comparable in order of magnitude to orbital Edelstein susceptibilities reported for other chiral conductors in the literature~\cite{doi:10.1021/acs.nanolett.7b04300, Salemi2022orbital}.

Having established both the characteristic energy dependence of $\chi_z^{L_z}$ and the normalization convention, we now examine how the orbital Edelstein susceptibility depends on the CNT geometry across the full set of metallic chiralities. In total, we consider all 25 metallic CNTs satisfying the condition $n + m \leq 20$, spanning diameters from $d = 3.58~\text{\AA}$ [the smallest CNT with $(4,1)$] to $d = 14.15~\text{\AA}$ [the largest CNT with $(17,2)$] and chiral angles from $\theta \approx 4^\circ$ to $\theta \approx 27^\circ$. Figure~\ref{fig:Fchimetal}(b) displays the Edelstein susceptibility $\chi_z^{L_z}$ as a function of CNT diameter at a fixed gate voltage $V_g = 0.25\,t$, at which all studied metallic CNTs remain within their first electronic subband (the smooth, single-channel regime discussed above). The resulting plot is reminiscent of the typical well-known Kataura representation~\cite{KATAURA19992555, Reich2004}, in which optical transition energies are mapped against diameter and reveal family-dependent patterns. Here, however, the mapped quantity is the orbital Edelstein susceptibility rather than an optical gap. At this energy, all nanotubes lie within the first electronic subband, corresponding to the smooth regime discussed previously. The results show that although the Edelstein susceptibility generally increases with nanotube diameter \cite{McEuen_muexp}, consistent with the expectation that larger circumferences support stronger azimuthal currents and hence larger orbital moments~\cite{McEuen_muexp, PhysRevLett.76.2121}, the data do not collapse onto a single universal curve \cite{KATAURA19992555, Reich2004}. Instead, the CNTs separate into two nearly linear branches, distinguished by their proximity to the zigzag or armchair limit, indicating that the diameter alone is insufficient to fully characterize the orbital magnetic response. For example, CNTs with small chiral angles (large $n/m$ ratio), such as $(13,1)$ and $(16,1)$, populate the upper branch and exhibit substantially larger susceptibilities, whereas CNTs closer to the armchair geometry, such as $(12,3)$ and $(11,8)$, fall on the lower branch. This bifurcation is particularly striking for the $(11,8)$ and $(16,1)$ pair, which share nearly identical diameters ($d \approx 12.9~\text{\AA}$) yet differ in the magnetic response of $\chi_z^{L_z}$ by almost a factor of two. The result demonstrates unambiguously that the orbital Edelstein response is not a simple geometric property of the nanotube cross-section but is instead controlled by the detailed subband structure, and, in particular, by the angular-momentum composition of the propagating channels, which is uniquely determined by the chiral wrapping vector $(n,m)$~\cite{RevModPhys.87.703, PhysRevB.92.075433}.

Figure~\ref{fig:Fchimetal}(c) maps the orbital Edelstein susceptibility in the 2D parameter space defined by nanotube diameter and chiral angle, providing an overview of how geometry and electronic structure jointly determine the orbital response. The magnitude of $\chi_z^{L_z}$ exhibits a pronounced dependence on both the nanotube diameter and chirality, growing broadly with increasing diameter, consistent with the behavior expected for a classical solenoid, where larger circumferential electronic orbits enhance the orbital magnetic response~\cite{PhysRevLett.76.2121, Ajiki1993}. However, chirality introduces substantial modulation of the susceptibility. Note that the largest values of $\chi_z^{L_z}$ are concentrated in the region of large diameters and small-to-intermediate chiral angles ($\theta \lesssim 15^\circ$), where the helical pitch of the wrapping vector most efficiently couples the longitudinal current to the azimuthal degree of freedom. Conversely, near-armchair CNTs ($\theta \to 30^\circ$) of comparable diameter exhibit systematically weaker orbital responses, reflecting the reduced helical winding of their Bloch states around the circumference. Nevertheless, even within specific chiral families, the evolution of the susceptibility $\chi_z^{L_z}$ is highly non-monotonic. For instance, along the $(n,1)$ branch, represented by nanotubes such as $(4,1)$, $(7,1)$, $(10,1)$, $(13,1)$, and $(16,1)$, increasing $n$ simultaneously enlarges the diameter and \emph{decreases} the chiral angle from $\theta_{(4,1)} \approx 10.9^\circ$ toward $\theta_{(16,1)} \approx 3.0^\circ$, driving the geometry progressively closer to the zigzag limit. Along this branch, $\chi_z^{L_z}$ increases monotonically, reflecting the combined effect of larger circumferential orbits and a progressively more favorable helical winding angle. We note, however, that when the gate voltage is increased beyond the first subband [as in Fig.~\ref{fig:Fchimetal}(a)], the susceptibility can exhibit irregular jumps that correlate with changes in the number and angular-momentum character of the subbands crossing the Fermi 
level~\cite{PhysRevB.92.075433, RevModPhys.87.703}.This behavior demonstrates that the orbital Edelstein response of metallic CNTs is governed not only by geometric parameters such as diameter and chiral angle, but also by the detailed electronic structure encoded in the chiral indices $(n,m)$ of the wrapping vector. In this sense, the orbital susceptibility constitutes a sensitive spectroscopic fingerprint of the nanotube's electronic chirality, analogous to the family-dependent optical transition patterns observed in Kataura plots~\cite{KATAURA19992555, Bachilo2002}.

To gain a more quantitative understanding of the chirality-dependent orbital response and to further support the picture of electronic currents circulating around the CNT circumference, we calculate the average ratio between the azimuthal current, $I_\phi$, responsible for generating the orbital angular momentum, and the axial current, $I_z$ that flows along the tube axis in response to the applied bias \cite{PhysRevB.57.9485, PhysRevB.80.235430}. To this end, we evaluate the local electric current at each atomic site with coordinates $(x,y,z)$ using Eq.~\eqref{eq:Iij}. The current is then decomposed into its Cartesian components $(I_x,I_y,I_z)$, from which we directly obtain the axial current $I_z$ and the circular current
\begin{equation}
I_\phi = \frac{-y I_x + x I_y}{\sqrt{x^2+y^2}} .
\end{equation}
The ratio $\langle I_\phi \rangle / \langle I_z \rangle$, averaged over all sites in the nanotube, thus quantifies the efficiency with which the chiral lattice structure deflects the longitudinal charge flow into a circulating azimuthal current. This quantity provides a direct, real-space measure of the helical winding of the current streamlines imposed by the chiral wrapping geometry and is intimately related to the orbital Edelstein susceptibility: a larger $I_\phi/I_z$ ratio implies a stronger chirality-mediated conversion of linear momentum into orbital angular momentum~\cite{PhysRevLett.76.2121, PhysRevB.78.233405}.

Figure~\ref{fig:Fchimetal}(d) compares the site-averaged current ratio $I_\phi/I_z$ for the $(16,1)$ and $(11,8)$ CNTs that correspond to two metallic chiralities that share nearly the same diameter ($d \approx 12.9~\text{\AA}$) yet belong to different chiral families. At low gate voltages  the ratio remains very small for both CNT systems, indicating that the electronic current flows predominantly along the nanotube axis with negligible azimuthal deflection. This is consistent with the approximately linear dispersion of the lowest subband, whose group velocity is directed predominantly along $z$~\cite{PhysRevLett.76.2121}. Thus, in this regime, the circumferential component of the current is negligible, leading to correspondingly small values of $\chi_z^{L_z}$. As the gate voltage $V_g$ increases the azimuthal component grows and the ratio saturates at markedly different plateaus: $I_\phi/I_z \approx 0.37$ for the near-zigzag $(16,1)$ CNT versus $I_\phi/I_z \approx 0.1$ for the near-armchair $(11,8)$ CNT. The threefold difference in the saturation value, despite identical diameters, directly reflects the stronger helical winding of the current streamlines in near-zigzag geometries, where the chiral angle $\theta$ is small and the wrapping vector makes a steep angle with respect to the circumferential direction~\cite{PhysRevB.78.233405, PhysRevB.60.13885}. This observation provides a transparent real-space explanation for the branch splitting seen in the Kataura-like plot of Fig.~\ref{fig:Fchimetal}(b): the upper branch (larger $\chi_z^{L_z}$) corresponds precisely to those chiralities that sustain a larger fraction of azimuthal current for a given longitudinal bias. Directly speaking, one has that these results reveal the emergence of a significant circumferential current component, particularly pronounced in the $(16,1)$ CNT.

Invoking the analogy again with a classical solenoid or a helical conducting wire, the quantity $\alpha_{\mathrm{eff}} = \tan^{-1}(I_\phi/I_z)$ may be interpreted as an effective winding angle of the electronic current relative to the nanotube axis~\cite{PhysRevLett.76.2121, PhysRevB.57.9485}. From the calculated current ratios, we obtain effective angles of approximately $\alpha_{\mathrm{eff}} \approx 19.8^\circ$ for the $(16,1)$ CNT and $\alpha_{\mathrm{eff}} \approx 5.5^\circ$ for the $(11,8)$ CNT, indicating a substantially tighter effective helical winding in the former case, with a nearly fourfold difference in the effective helical pitch of the current flow despite the identical tube diameters. Notably, the geometric chiral angles of these two tubes are $\theta_{(16,1)} \approx 3.0^\circ$ and $\theta_{(11,8)} \approx 26.3^\circ$, respectively. The fact that the near-zigzag tube $(16,1)$ exhibits a \emph{larger} effective current winding angle than its geometric chiral angle, while the near-armchair tube $(11,8)$ shows the opposite trend, underscores that $\alpha_{\mathrm{eff}}$ is not simply a proxy for $\theta$ but rather encodes the collective angular-momentum composition of all occupied transport channels. It is important to emphasize that these values for $\alpha_{\mathrm{eff}}$ represent site-averaged quantities extracted from the full bond-current distribution. On the discrete hexagonal lattice, the current is carried along the three nearest-neighbor bond directions, and the zigzag-oriented bonds, which make the smallest angle with the CNT axis, typically sustain the largest current density~\cite{PhysRevB.60.13885}. The electronic current flow is therefore not strictly confined to a single helical path but is distributed among all bond orientations with weights that depend on the specific subband occupation at a given $V_g$. Consequently, the extracted angle $\alpha_{\mathrm{eff}}$ should be regarded as an effective average angle between the current flow and the nanotube axis, rather than the geometrical chiral angle directly determined by the $(n,m)$ indices. This distinction between the structural chiral angle $\theta$ and the transport-derived winding angle $\alpha_{\mathrm{eff}}$ may partially explain the absence of a simple monotonic relationship between the calculated $\chi_z^{L_z}$ and the nanotube's geometric parameters: the orbital response is ultimately governed by the quantum-mechanical current texture, which depends on the detailed subband filling and cannot be reduced to a single geometric descriptor.

\subsection{Semiconducting Carbon Nanotubes }

\begin{center}
\begin{figure}[!t]
\scalebox{0.85}{\includegraphics{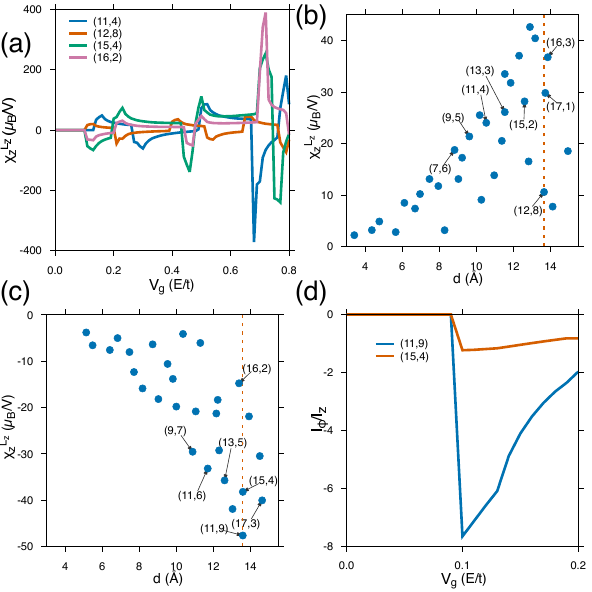}}
\caption{\textbf{Orbital Edelstein susceptibility of semiconducting chiral CNTs.} All values of $\chi_z^{L_z}$ are normalized by the respective unit-cell length $|\mathbf{T}|$. (a) Orbital Edelstein susceptibility, $\chi_z^{L_z}$, as a function of gate voltage for representative semiconducting CNTs, illustrating the $\sim\!\sqrt{E}$ onset, the intra-subband sign change, and the oscillatory multi-subband regime. (b) Dependence of the normalized orbital Edelstein susceptibility on nanotube diameter for all semiconducting CNTs exhibiting positive orbital response ($\chi_z^{L_z} > 0$). For each nanotube, the gate voltage was chosen at the center of the first conducting subband. The data segregates into distinct family-dependent branches analogous to a Kataura plot. (c) Same as panel (b), but for semiconducting CNTs with $\chi_z^{L_z}<0$, confirming that the branch structure persists irrespective of the sign of the orbital response. (d) Ratio between the azimuthal and axial current components, $I_{\phi}/I_z$, for the $(11,9)$ and $(15,4)$ CNTs, both having a diameter of $d = 13.58~\text{\AA}$, but belong to different chiral families, demonstrating the dominant role of the wrapping geometry over the tube radius in determining the circumferential current magnitude.}
\label{fig:Fchisemi} 
\end{figure}
\end{center}

We now turn to semiconducting CNTs, which satisfy $(n - m)\,\mathrm{mod}\,3 \neq 0$~\cite{bookSaito, RevModPhys.79.677}. Following the same protocol as for the metallic case, we shall characterize the non-equilibrium orbital magnetic response of semiconducting CNTs with chiral indices within the sector $0 \leq m \leq n$ and the constraint \(n+m \leq 20\), as discussed previously, spanning from the narrowest CNT $(2,1)$ to the widest $(18,2)$ and yielding a total of 61 distinct semiconducting chiralities of studied CNTs. Representative energy-dependent lineshapes of the orbital Edelstein susceptibility are shown in Fig.~\ref{fig:Fchisemi}(a). Compared with metallic CNTs, the semiconducting response exhibits several qualitatively distinct features that merit individual discussion. First, semiconducting CNTs exhibit a non-equilibrium orbital magnetic response substantially larger than their metallic counterparts of comparable diameter. This enhancement can be traced to the van Hove singularity at the subband edge: the divergent density of states characteristic of one-dimensional systems amplifies the orbital response near the band onset, an effect absent in metallic tubes whose lowest subbands disperse linearly through the charge-neutrality point~\cite{RevModPhys.79.677, Mintmire1998}. Second, the energy dependence within the first conducting subband follows a $\sim\!\sqrt{E - E_{\mathrm{gap}}/2}$ behavior, reflecting the parabolic dispersion near the subband minimum and the associated one-dimensional density of states $\rho(E) \propto (E-E_{\mathrm{edge}})^{-1/2}$~\cite{bookSaito}. Finally, at higher energies where multiple subbands become available for conduction and thus contribute to transport, the susceptibility oscillates between positive and negative values as successive channels with different crystal angular momenta are activated; a multi-channel interference effect analogous to that observed in the metallic case~\cite{PhysRevB.78.233405}.

To systematically map the diameter dependence of the orbital Edelstein susceptibility for all studied semiconducting CNTs, \textit{i.e.}, across all 61 semiconducting chiralities, we separate the systems according to the sign of $\chi_z^{L_z}$. We note that, within the nearest-neighbor tight-binding model, the Hamiltonian possesses an exact electron-hole symmetry, implying $\chi_z^{L_z}(-V_g) = -\chi_z^{L_z}(V_g)$. Consequently, the orbital response in the hole-doped regime ($V_g < 0$) is fully determined by the electron-doped results presented here, and we restrict the discussion to $V_g > 0$ for clarity. The sign is uniquely determined by the family index: CNTs satisfying $(n - m)\,\mathrm{mod}\,3 = 1$ (type-I semiconductors) yield $\chi_z^{L_z} > 0$, while those with $(n - m)\,\mathrm{mod}\,3 = 2$ (type-II) yield $\chi_z^{L_z} < 0$~\cite{Reich2004, KATAURA19992555}. This classification mirrors the well-known family splitting observed in optical spectroscopy of CNTs~\cite{Bachilo2002, Reich2004}, and its appearance in the orbital Edelstein response constitutes a direct manifestation of the distinct angular-momentum textures characterizing the two semiconductor families. For a consistent comparison, the susceptibility is evaluated at the energy corresponding to the center of the first conducting subband for each chirality. The results are displayed in Figs.~\ref{fig:Fchisemi}(b) and \ref{fig:Fchisemi}(c) for the positive and negative branches of the susceptibility, respectively. Similar to the metallic case, the data points do not collapse onto a single universal function, confirming that the CNT diameter alone is insufficient to parameterize the orbital response. Nevertheless, a clear overall linear increase of $|\chi_z^{L_z}|$ with nanotube diameter is observed, consistent with the semiclassical expectation that larger circumferential orbits generate proportionally stronger orbital moments~\cite{PhysRevLett.76.2121, Ajiki1993}. In addition, the formation of distinct branches becomes more evident in semiconducting CNTs, likely due to the larger number of data points available in this case. This branch formation in the metallic case, owing to the larger number of chiralities available, is the hallmark of the so-called ``family behavior'' of CNTs, in which tubes related by the transformation $(n,m) \to (n+2,\, m-1)$ share similar electronic properties and form a single branch in Kataura-type plots~\cite{KATAURA19992555, Reich2004}.

The family structure is particularly transparent for the positive-susceptibility values; the evolution of \(\chi_z^{L_z}\) along one of the observed branches can be traced to the sequence of tubes \((7,6)\), \((9,5)\), \((11,4)\), \((13,3)\), \((15,2)\), and \((17,1)\), along which $n$ increases by 2 while $m$ decreases by 1 at each step. This progression simultaneously enlarges the diameter and rotates the chiral angle from the near-armchair limit [with chiral angle $\theta_{(7,6)} = 27.46^\circ$] toward the near-zigzag limit [with chiral angle $\theta_{(17,1)} = 2.83^\circ$]. The monotonic increase of $\chi_z^{L_z}$ along this branch demonstrates that, within a given family, the orbital response is maximized for chiralities closest to the zigzag geometry, precisely those for which the helical pitch of the wrapping vector most effectively deflects the longitudinal current into azimuthal circulation, as discussed in the metallic case. This systematic trend provides a practical design rule: for a target diameter, the semiconducting chirality with the smallest chiral angle within a given family will exhibit the strongest orbital Edelstein effect. A similar behavior is observed for negative susceptibility values, corresponding to the branch defined by the CNTs \((9,7)\), \((11,6)\), \((13,5)\), \((15,4)\), and \((17,3)\). In this case, the evolution from a nearly armchair geometry to a near-zigzag geometry occurs from the \((9,7)\) CNT with \(\theta = 25.87^{\circ}\), to the \((17,3)\) CNT with \(\theta = 7.99^{\circ}\).

The same family-dependent trend is observed across all identified branches, and its most consequential implication becomes apparent when different branches intersect in the diameter axis. Because each family follows its own approximately linear trajectory in the $\chi_z^{L_z}$--diameter plane, tubes belonging to distinct families can share nearly the same diameter yet exhibit markedly different orbital responses, \textit{i.e.} when only the nanotube diameter is considered, different branches can intersect, resulting in distinct values of $\chi_z^{L_z}$ for nanotubes with nearly identical diameters but different chiral angles. This situation is illustrated in Fig.~\ref{fig:Fchisemi}(c) for the $(15,4)$ and $(11,9)$ CNTs, which possess the same diameter $d = 13.58~\text{\AA}$ but belong to different chiral families [$(n-m)\,\mathrm{mod}\,3 = 2$ in both cases, yet with distinct family indices $2n + m$]. Despite their identical cross-sections, the $(11,9)$ tube (with larger chiral angle $\theta = 26.69^\circ$) exhibits a significantly larger susceptibility $|\chi_z^{L_z}|$ than the $(15,4)$ tube ($\theta = 11.5^\circ$), as depicted in Fig.~\ref{fig:Fchisemi}(d). At first glance, this appears to contradict the within-family rule established above---namely, that smaller chiral angles yield larger susceptibilities. The resolution lies in the fact that the two tubes reside on \emph{different} branches: the $(11,9)$ tube sits on a branch whose overall intercept in the $\chi_z^{L_z}$--diameter plane is higher, reflecting a more favorable angular-momentum composition of its first conducting subband~\cite{PhysRevB.92.075433}. This example underscores that the orbital Edelstein response is a genuinely two-parameter quantity---determined jointly by the diameter \emph{and} the family membership encoded in the chiral vector---and cannot be predicted from any single geometric descriptor alone.

The larger susceptibility is also reflected in a substantially larger azimuthal-to-axial current ratio $I_{\phi}/I_z$ for the $(11,9)$ CNT, as shown in Fig.~\ref{fig:Fchisemi}(d). In contrast to the metallic case, where $I_\phi/I_z$ saturates to a roughly constant plateau within the first subband, making it difficult to straightforwardly associate its value with an effective winding angle of a solenoid, the semiconducting ratio displays a pronounced energy dependence that precludes such a simple geometric analogy. Physically, this behavior can be understood in terms of the dispersion relation near the subband edge. At the onset of conduction, the longitudinal group velocity $v_z = \partial E/\partial k_z$ vanishes (the band bottom is a turning point), while the azimuthal velocity component---set by the transverse quantization condition and the angular-momentum quantum number remains finite~\cite{PhysRevB.92.075433,bookSaito}. Consequently, the current flow near the band edge is dominated by its circumferential component, yielding a large $I_\phi/I_z$. As the Fermi level moves deeper into the subband, $v_z$ grows (the dispersion becomes increasingly linear) while the azimuthal component, which is bounded by the circumferential quantization, saturates or even decreases. The net effect is a monotonic reduction of $I_\phi/I_z$ with increasing energy within the subband. This energy-dependent redistribution of current between the axial and azimuthal directions is also the microscopic origin of the intra-subband sign change of $\chi_z^{L_z}$ noted earlier: as the balance between $I_\phi$ and $I_z$ shifts, the net orbital moment can reverse its orientation.

\subsection{Metallic contacts}
\label{sec:metalcontact}

\begin{center}
\begin{figure}[!t]
\scalebox{0.85}{\includegraphics{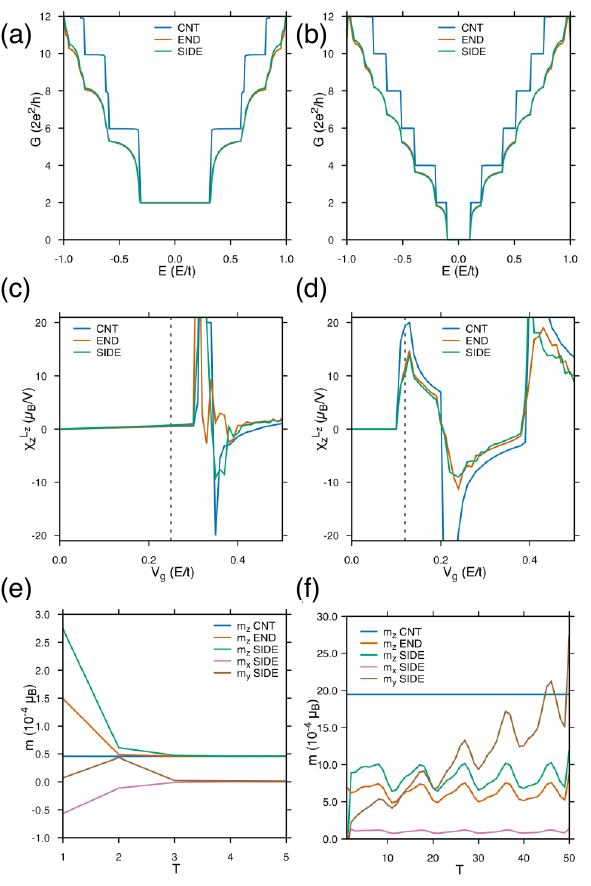}}
\caption{\textbf{Influence of the contact geometry on charge transport and orbital magnetization.} (a,\,b)~Two-terminal conductance of the metallic $(11,8)$ and semiconducting $(12,8)$ CNTs, respectively, computed with three different contact configurations: ideal semi-infinite CNT leads (CNT), wide-band end-contact geometry (END), and wide-band side-contact geometry (SIDE). The near-identical traces for END and SIDE confirm that total charge transmission is insensitive to the contact geometry within the wide-band approximation. (c,\,d)~Orbital Edelstein susceptibility $\chi_z^{L_z}$ (normalized by the unit-cell length $|\mathbf{T}|$) as a function of gate voltage for the same $(11,8)$ and $(12,8)$ CNTs and same contact configurations. For the metallic tube, $\chi_z^{L_z}$ is contact-independent in the first subband; for the semiconducting tube, metallic contacts systematically reduce the susceptibility. (e,\,f)~Unit-cell-averaged components of the spatially resolved orbital magnetic moment $\mathbf{m} = (m_x, m_y, m_z)$ expressed in units of the Bohr magneton ($\mu_B$) along the nanotube axis for the $(11,8)$ CNT at $V_g = 0.25\,t$ [indicated by the dashed line in (c)] and the $(12,8)$ CNT at $V_g = 0.12\,t$ [center of the first conductance plateau, indicated by the dashed line in (d)]. The metallic CNT rapidly recovers its bulk orbital moment within a few unit cells, whereas the semiconducting tube exhibits persistent oscillations indicative of mode beating between channels with different crystal angular momenta.}
\label{fig:endside} 
\end{figure}
\end{center}

The results presented so far have shown that a longitudinal charge current flowing through a chiral CNT, whether metallic or semiconducting, generates a measurable orbital magnetization whose magnitude and sign are governed by the chiral wrapping vector. The large macroscopic susceptibility indicates that the current acquires a significant azimuthal component, effectively forming microscopic solenoidal current patterns, \textit{i.e.} the chiral lattice efficiently deflects the longitudinal flow into circumferential circulation, effectively turning the nanotube into a microscopic current solenoid. However, up to this point, all calculations presented thus far employed ideal semi-infinite CNT leads on both sides of the device [Fig.~\ref{fig:fig1}(a)]. In this configuration, the current injected from the lead already carries the full angular-momentum texture of the nanotube's propagating Bloch states because the lead and the scattering region share the same electronic structure. Consequently, the measured orbital response reflects not only the intrinsic chirality-induced selectivity of the finite nanotube segment but also the orbital content pre-encoded in the incoming wave functions~\cite{Datta1995, Nemec2006}. To disentangle these two contributions and to isolate the \emph{intrinsic}  angular momentum selectivity of the CNTs, we now replace the left semi-infinite ideal CNT electrode by a featureless wide-band metallic contact [Figs.~\ref{fig:fig1}(b)--\ref{fig:fig1}(c)]. The wide-band self-energy $\Sigma_L = -i\,|t|$ injects electrons with a flat spectral density and, crucially, without any preferential angular-momentum polarization and without a predefined angular-momentum distribution, it couples uniformly to all orbital channels at the contact interface~\cite{Bahamon, Datta1995}, enabling a more direct evaluation of the CNTs' angular-momentum selectivity. Any orbital magnetization that develops downstream of the contact must therefore originate entirely from the chirality-induced redistribution of current among the nanotube's angular-momentum channels, providing a stringent test of the CNT's intrinsic orbital selectivity. We emphasize that the ideal semi-infinite CNT right lead shares the same Hamiltonian as the device region, so that the right interface is structurally seamless and the reflection coefficient vanishes identically. All scattering and mode-mixing phenomena reported below therefore originate exclusively from the left metallic contact. This asymmetric configuration is deliberately chosen to provide the cleanest probe of the CNT's intrinsic angular-momentum imprinting: a symmetric setup with metallic contacts on both sides would introduce additional Fabry--P\'erot-like resonances from back-scattering at the right interface, whose device-length-dependent oscillation period would superimpose on the intrinsic forward-propagating mode beating and complicate its identification. In Appendix~\ref{sec.appB}, we present results for the metallic $(11,8)$ CNT with metallic leads at both the left and right ends, with both contacts in the end-contact configuration.

It is well established that the nature of the metal--nanotube interface profoundly influences the transport properties of CNT-based devices: the contact material, its work function, the bonding geometry, and the spatial extent of the coupling region all affect transmission coefficients and channel mixing~\cite{Nemec2006, PhysRevLett.90.106801, Nemec, Tersoff1999, Chen2005contact}. A full \textit{ab initio} treatment of the contact is beyond the scope of the present work; our objective here is not to model a specific metal--CNT junction but rather to inject current that is \emph{angularly unpolarized}, \textit{i.e.}. To this end, we adopt the wide-band approximation, as discussed earlier, and consider two experimentally motivated contact geometries that differ in the subset of atoms coupled to the metallic reservoir. The first is the \textit{end-contacted} configuration [Fig.~\ref{fig:fig1}(b)], where the self-energy acts on all atoms at the left edge of the CNT, representing a scenario in which the tube terminates at a planar electrode and current is injected uniformly across the full cross-section. The second is the \textit{side-contacted} configuration [Fig.~\ref{fig:fig1}(c)], where only atoms satisfying $y \geq 0$ within the leftmost unit cell are coupled to the lead, mimicking a nanotube deposited on a metallic substrate where the coupling is restricted to the lower hemisphere of the tube~\cite{Tersoff1999}. The key distinction between these two geometries is that the end contact preserves the discrete rotational symmetry $C_N$ of the nanotube cross-section at the injection plane, whereas the side contact explicitly breaks it, thereby selectively addressing only a subset of angular-momentum channels at the interface.

Before analyzing the orbital magnetization profile and calculating the angular momentum susceptibility, it is instructive to first characterize the charge injection efficiency of the metallic contacts. This is achieved by computing the two-terminal conductance $G = (2e^2/h)\,T(E)$, where $T(E)$ is the energy-resolved transmission probability function obtained within the Green's function formalism~\cite{Bahamon, Datta1995}. Although we only present representative results for a metallic $(11,8)$ and a semiconducting $(12,8)$ CNT, shown in Figs.~\ref{fig:endside}(a)--\ref{fig:endside}(b), the qualitative main conclusions discussed below hold and are fully extended for all chiralities in our dataset to all CNTs studied. Two key observations emerge. First, from the standpoint of total charge transmission, the end-contacted and side-contacted geometries yield nearly indistinguishable conductance traces, \textit{i.e.}, essentially no difference in charge injection between the two. This insensitivity is expected within the wide-band limit, where the self-energy $\Sigma_L = -i|t|$ provides a broadening comparable to the bandwidth, ensuring that all propagating modes at a given energy are coupled to the reservoir with similar strength regardless of the precise spatial distribution of the contact atoms~\cite{Nemec2006, Bahamon}. Second, the injection efficiency depends on the number of available channels: for metallic CNTs in the lowest subband (two linearly dispersing modes), the wide-band contact achieves near-perfect transmission ($T \approx 2$), consistent with the well-known result that a single-mode quantum wire coupled to a broad-band reservoir transmits with unit probability~\cite{Datta1995}. At higher energies, where additional subbands open, or in semiconducting CNTs whose first subband has a parabolic dispersion and a smaller group velocity at the band edge, the mode-matching between the featureless metallic reservoir and the structured nanotube density of states becomes less favorable, resulting in reduced transmission~\cite{Bahamon, Nemec}. Importantly, although the total charge conductance is largely geometry-independent, the \emph{orbital} content of the injected current, which determines the downstream magnetization profile, will prove to be sensitive to the contact geometry, as we demonstrate below.

Having established that the wide-band contacts inject charge without imprinting spurious spectral features on the conductance, we now examine how they affect the orbital Edelstein susceptibility. Figure~\ref{fig:endside}(c) displays the Edelstein susceptibility $\chi_z^{L_z}$ for the metallic $(11,8)$ CNT computed with ideal CNT leads, end contacts, and side contacts. Within the energy window of the first conducting subband, where a single pair of linearly dispersing modes carries the current, the susceptibility is essentially independent of the contact type. This robustness is physically expected: when only one propagating channel per valley is available, the angular-momentum composition of the transmitted state is uniquely fixed by the nanotube band structure, leaving no room for the contact geometry to alter the orbital content~\cite{PhysRevB.92.075433}. At higher energies, where multiple subbands with distinct crystal angular momenta become active, the picture changes qualitatively. The metallic contact, which injects into all available channels with comparable weight, can produce inter-channel interference that manifests as oscillations and locally enhanced values of $\chi_z^{L_z}$. Nonetheless, near the center of each higher conducting subband, the susceptibility $\chi_z^{L_z}$ is systematically reduced relative to the ideal-lead case. This behavior can be understood from the conductance curves. Note that this reduction can be correlated directly with the decreased transmission observed in the conductance: imperfect mode-matching at higher energies suppresses the contribution of certain angular-momentum channels, thereby diminishing the net orbital moment~\cite{Bahamon, Nemec}. The effect is even more pronounced for the semiconducting $(12,8)$ CNT [Fig.~\ref{fig:endside}(d)], where the transmission deficit extends across the entire first subband owing to the parabolic dispersion and reduced group velocity at the band edge. Consequently, $\chi_z^{L_z}$ is systematically smaller with metallic contacts throughout the semiconducting energy window. These observations highlight an important practical consideration: while the \emph{qualitative} chirality-dependent features of the orbital response (sign, family structure, branch splitting) are robust against the contact model, the \emph{quantitative} magnitude of $\chi_z^{L_z}$ is sensitive to the injection efficiency and must be assessed with realistic contact descriptions for device-level predictions.

To gain insight into the spatial mechanism by which the injected current acquires orbital angular momentum, we resolve the average orbital magnetic moment $\mathbf{m} = (m_x, m_y, m_z)$ for each unit cell along the CNT device axis using either CNT or metallic left contacts. Unit cell 1 corresponds to the first cell of the device adjacent to the left contact, i.e., the point where the unpolarized current is injected into the nanotube. Figure~\ref{fig:endside}(e) displays the spatial profile of $\mathbf{m}$ for the metallic $(11,8)$ CNT at $V_g = 0.25\,t$. For the CNT-contacted device, the axial component $m_z$ is spatially uniform and reaches an approximately constant value of $m_z \approx 0.46\times10^{-4}\,\mu_B$, as expected for a translationally invariant system. When a wide-band metallic contact replaces the left lead, the situation near the injection point changes qualitatively: the incoming current carries no preferential angular-momentum polarization, and the nanotube must ``imprint'' its intrinsic orbital texture onto the transmitted electrons through the chiral potential landscape. Remarkably, this imprinting occurs over an extremely short length scale. Within only three to four unit cells from the contact, corresponding to roughly $3$--$5$~nm for the $(11,8)$ tube, the axial moment $m_z$ converges to the same bulk value obtained with ideal leads, indicating that the metallic CNT rapidly projects the injected state onto its eigenchannels and establishes the equilibrium angular-momentum distribution. The end-contact geometry achieves convergence slightly faster ($\sim\,3$ unit cells) than the side contact ($\sim\,4$ unit cells), consistent with the fact that the end contact preserves the $C_N$ rotational symmetry and therefore does not excite transverse ($m_x$, $m_y$) components that must subsequently relax. This rapid equilibration stands in marked contrast to wave-packet propagation studies on helical structures~\cite{doi:10.1021/acs.jctc.5c01410,10.1063/5.0156491}, where the conversion efficiency between linear and angular momentum depends sensitively on the helical pitch and can require many turns to reach steady state. In the present steady-state transport framework, the strong coupling between all propagating channels mediated by the chiral lattice ensures that the angular-momentum redistribution is accomplished within a few lattice periods. In the immediate vicinity of the contact, non-negligible transverse components $m_x$ and $m_y$ appear, a direct signature of broken translational symmetry at the lead-nanotube interface. These transverse moments decay rapidly toward zero within the same few unit cells over which $m_z$ converges, confirming that the contact-induced perturbation is confined to a narrow boundary layer and does not propagate into the bulk of the nanotube.

The situation is qualitatively different for the semiconducting $(12,8)$ CNT at $V_g = 0.12\,t$, shown in Fig.~\ref{fig:endside}(f). Based on the large Edelstein susceptibilities and strong azimuthal currents established for semiconducting tubes in the preceding sections, one might expect the orbital moment to converge rapidly to a well-defined bulk value, perhaps reduced in magnitude relative to the ideal-lead case due to the lower transmission, but otherwise spatially uniform. Instead, all components of $\mathbf{m}$ exhibit pronounced and persistent spatial oscillations that extend throughout the device, with amplitudes and means that depend markedly on the contact geometry. This oscillatory behavior can be understood as a consequence of quantum beating between the multiple transport channels available in the semiconducting subband. Unlike the metallic case, where a single pair of linearly dispersing modes dominates at low energy and uniquely fixes the angular-momentum content, the semiconducting first subband generally supports modes with different longitudinal wave vectors $k_z$ at the same energy (arising from the two valleys $K$ and $K'$, which carry distinct  angular momenta). When these modes are coherently excited by the metallic contact, which, being featureless, populates both valleys with comparable amplitude, their superposition produces a spatially modulated interference pattern in the bond-current distribution, and hence in the orbital moment~\cite{PhysRevB.91.235442, PhysRevB.93.195442}.  Quantitatively, the $m_z$ component oscillates in the range $(5\text{--}7.5)\times10^{-4}\,\mu_B$ for the end-contact geometry and $(7\text{--}10)\times10^{-4}\,\mu_B$ for the side contact. The systematically larger mean value in the side-contact case suggests that breaking the $C_N$ symmetry at the interface preferentially populates the angular-momentum channel with the larger orbital moment, effectively biasing the injection toward one valley. The transverse component $m_x$ remains comparatively small ($\sim 1\times10^{-4}\,\mu_B$ for the side contact), indicating that the oscillations are predominantly axial in character and arise from interference in the azimuthal current rather than from a precession of the moment direction.

The most striking feature, however, is the behavior of the transverse component 
$m_y$: rather than remaining small or decaying away from the contact, it grows monotonically with distance along the nanotube and develops large-amplitude oscillations, eventually reaching, and even exceeding, the magnitude of the axial moment $m_z$ obtained in the ideal CNT-contacted geometry. This progressive buildup of a transverse orbital moment, superimposed with a well-defined oscillation period, is indicative of coherent mode beating between transport channels carrying different  angular momenta~\cite{PhysRevB.82.115320, PhysRevB.76.205315, PhysRevB.91.235442}. Physically, the side-contact geometry, by breaking the discrete rotational symmetry $C_N$ at the injection plane, does not project the incoming electrons onto a single angular-momentum eigenstate of the nanotube. Instead, it promotes a coherent superposition of modes with distinct eigenvalues ($\mu$) --- interference between multiple propagating modes --- each propagating with a different longitudinal wave vector $k_z(\mu)$. As these modes accumulate a relative phase $\Delta\phi = [k_z(\mu) - k_z(\mu')]\,z$ along the tube axis, the interference pattern rotates the direction of the net orbital moment away from the $z$-axis, generating a growing transverse component whose envelope is modulated by the beating wavelength $\lambda_{\mathrm{beat}} = 2\pi/|k_z(\mu) - k_z(\mu')|$. The resulting orbital magnetization thus acquires a pronounced transverse component that is entirely absent in the ideal-lead or end-contact configurations, demonstrating that the contact geometry can qualitatively reshape the spatial texture of the orbital moment in semiconducting CNTs. From a practical standpoint, this finding implies that side-contacted semiconducting CNTs can generate transverse orbital fields comparable in magnitude to the longitudinal Edelstein response, opening a route toward engineering the \emph{direction}, not merely the magnitude, of the current-induced orbital magnetization through contact design.

\section{Angular Momentum Filtering}
\label{sec:filtering}

In Section~\ref{sec:model}, we described the CNT geometry within the translational scheme defined by the pair $[\mathbf{C}_h, \mathbf{T}]$, which naturally introduces the quantum numbers $[k_\parallel, k_\perp]$, the longitudinal and transverse components of the Bloch wave vector, to label the electronic states~\cite{bookSaito, Reich2004, BARROS2006261}. Imposing periodic boundary conditions around the CNT circumference leads to the quantization of the transverse wave vector as
\begin{equation}
k_\perp = \frac{2\pi}{|\mathbf{C}_h|}\,n_c\,,
\end{equation}
where the integer $n_c$ is the pseudo-angular momentum quantum number that indexes the allowed subbands~\cite{PhysRevB.103.L100409}. An alternative and physically more transparent description is provided by the \emph{helical} (or line-group) representation of the nanotube, whose unit cell contains only two carbon atoms regardless of the chirality~\cite{Mintmire, MINTMIRE1995893, BARROS2006261}. In this scheme, this quantity is related to the crystal angular momentum $\mathfrak{m}$ through \( n_c=\mathfrak{m}\ \mathrm{mod}\ \mathfrak{n},\) that connects the two quantum numbers and with $\mathfrak{n}=\mathrm{gcd}(n,m)$ being the greatest common divisor of the CNT chiral indices $n$ and $m$ \cite{Lunde}, \textit{i.e.} the order of the principal rotational axis $C_N$ of the nanotube~\cite{PhysRevB.93.195442}. This relationship emerges when the electronic bands obtained in the helical-Brillouin-zone representation are back-folded into the translational-zone representation \cite{Lunde}. The main advantage of the helical representation is that its unit cell contains only two atoms and that each band is characterized by a well-defined crystal angular momentum quantum number $\mathfrak{m}\in\{0,1,\ldots,\mathfrak{n}-1\}$ \cite{Mintmire, MINTMIRE1995893}. This quantum number provides a natural framework for classifying transport in metallic and semiconducting CNTs based on the angular-momentum character of their low-energy electronic states \cite{PhysRevB.93.195442, Lunde}. Moreover, note that the crystal angular momentum $\mathfrak{m}$ is a rigorously conserved quantity under the $C_N$ rotational symmetry, allowing to classify states belonging to different $\mathfrak{m}$ sectors that propagate independently in a pristine, infinitely long tube and cannot scatter into one another~\cite{PhysRevB.93.195442, PhysRevB.92.075433, Marganska2019, BARROS2006261}. This conservation law is the basis of the angular-momentum filtering mechanism explored in the remainder of this section: by engineering the angular-momentum content of the injected state, one can exploit the nanotube as a selective transmission channel for a specific $\mathfrak{m}$ quantum number.

Within the helical classification, metallic CNTs are further subdivided into two symmetry classes~\cite{Lunde}. A metallic CNTs belongs to the \emph{armchair-class} if $(n - m)/\mathfrak{n}\;\mathrm{mod}\;3 = 0$; otherwise it is of the \emph{zigzag-class}. The distinction has direct consequences for the angular-momentum content of the low-energy transport channels: in the armchair class, both bands crossing the Fermi level ($E_F = 0$) are characterized by a crystal angular momentum $\mathfrak{m} = 0$, whereas in the zigzag class the two degenerate Fermi-level bands possess distinct crystal angular momenta $\mathfrak{m}_a = (2n+m)/3\;\mathrm{mod}\;\mathfrak{n}$ and $\mathfrak{m}_b = (2m+n)/3\;\mathrm{mod}\;\mathfrak{n}$~\cite{Lunde, PhysRevB.93.195442}. This means that zigzag-class metallic CNTs naturally support two co-propagating channels distinguishable solely by their angular momentum, a prerequisite for selective filtering. Motivated by this observation, we design a numerical experiment to probe the angular-momentum filtering capability of chiral CNTs. The key idea is to inject electrons that carry a \emph{well-defined} orbital angular texture and momentum $l$, to calculate the conductance, and to monitor whether the nanotube selectively transmits only those channels whose crystal angular momentum $\mathfrak{m}$ matches the injected value, in order to probe the orbital-angular-momentum filtering properties of CNTs. To construct such an orbitally textured injection of electronic states, we start from the wide-band contacts in the end-contact self-energy geometry and introduce azimuthal phase correlations between the contact sites by projecting the contact Green's function onto a state of definite orbital angular momentum $l$. Specifically, the resulting projected left contact Green's function reads
\begin{equation}
G_{i,j}(l) = \sum_{i=1}^{N_{\phi}}\sum_{j=1}^{N_{\phi}} \left(e^{-i l \phi_j}/\sqrt{N_{\phi}}\right) G_{\mathrm{WB}} \left(e^{i l \phi_i}/\sqrt{N_{\phi}}\right),
\label{eq:Gl_proj}
\end{equation}
where $N_\phi$ is the number of CNT atoms in the contact cross-section (\textit{i.e.}, the atoms around the circumference that are coupled to the left lead), $\phi_i$ denotes the azimuthal angle of site $i$, and $l$ is the integer specifying the injected orbital angular momentum. This construction is the discrete-lattice analog of selecting a single azimuthal Fourier component $e^{il\phi}$ from a cylindrically symmetric source: the phase factors $e^{\pm il\phi}$ impose a definite winding number on the injected wave front, analogous to the generation of optical vortex beams carrying orbital angular momentum $l\hbar$~\cite{Allen1992, BahamonChV3}. By varying $l$ and computing the resulting transmission, we can map out the angular-momentum-resolved transmission spectrum of the nanotube and quantify its filtering selectivity.

\begin{center}
\begin{figure}[!t]
\scalebox{0.85}{\includegraphics{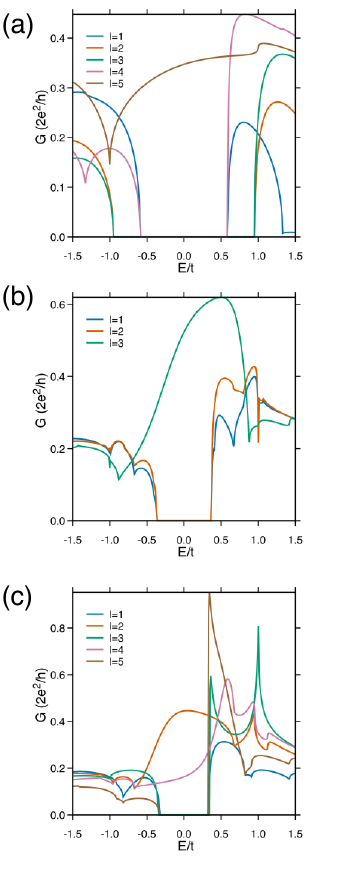}}
\caption{\textbf{Angular-momentum filtering in metallic carbon nanotubes.} Two-terminal conductance as a function of energy for electrons injected with different orbital angular momenta $l$ [Eq.~\eqref{eq:Gl_proj}] into (a)~the achiral armchair $(5,5)$ CNT ($\mathfrak{n} = 5$, armchair class, low-energy bands with $\mathfrak{m} = 0$), (b)~the chiral armchair-class $(12,3)$ CNT ($\mathfrak{n} = 3$, $\mathfrak{m} = 0$), and (c)~the chiral zigzag-class $(12,6)$ CNT ($\mathfrak{n} = 6$, low-energy bands with $\mathfrak{m}_a = 4$ and $\mathfrak{m}_b = 2$). Angular-momentum gaps---energy windows of vanishing transmission---appear whenever the injected $l$ does not satisfy the selection rule $l \equiv \mathfrak{m}\pmod{\mathfrak{n}}$. In (a) and (b), only $l \equiv 0\pmod{\mathfrak{n}}$ transmits near $E = 0$; in (c), both $l = 2$ and $l = 4$ are transmitted, each coupling selectively to a single angular-momentum channel without cross-talk.}
\label{fig:Filter} 
\end{figure}
\end{center}

To test the proposed angular-momentum injection strategy, we first apply this scheme to the achiral armchair $(5,5)$ CNT, for which the electronic band structure is fully characterized by crystal angular momenta $\mathfrak{m} \in \{0, 1, 2, 3, 4\}$ (since $\mathfrak{n} = \gcd(5,5) = 5$). Figure~\ref{fig:Filter}(a) displays the two-terminal conductance as a function of energy for several values of the injected angular momentum $l$. The first point to note is that the armchair CNT acts as an angular-momentum filter, exhibiting transmission gaps that depend on the injected orbital texture. For each value of $l$, transmission is permitted only within energy windows where propagating subbands with matching crystal angular momentum exist, while \emph{angular-momentum gaps}---energy regions devoid of states with the appropriate $\mathfrak{m}$---appear as zero-transmission windows. The filtering mechanism is most transparent near the Fermi level ($E = 0$), where the two linearly dispersing bands of the $(5,5)$ tube both carry $\mathfrak{m} = 0$ (armchair class). Consequently, only injected states satisfying the selection rule $l \equiv \mathfrak{m} \pmod{\mathfrak{n}}$, with $\mathfrak{m} = 0$, can couple to these low-energy channels. For $\mathfrak{n} = 5$, this condition is met by $l = 0, 5, 10, \ldots$\,---\textit{i.e.}, by any $l$ that is a multiple of $\mathfrak{n}$. Indeed, injection with $l = 5$ yields a finite transmission around $E = 0$, while all other values of $l$ tested produce a transmission gap at the Fermi level. The equivalence between $l = 5$ and $l = 10$ (not shown) confirms that the filtering condition is periodic in $l$ with period $\mathfrak{n}$, as expected from the discrete rotational symmetry $C_5$ of the nanotube cross-section~\cite{PhysRevB.93.195442, Lunde}. This periodicity is the angular-momentum analog of Bloch's theorem in the azimuthal direction: just as $k_\perp$ is defined modulo a reciprocal lattice vector, the effective angular momentum relevant for transmission is defined modulo $\mathfrak{n}$.

Secondly, we emphasize that the primary purpose of the projected contact is not to model a specific experimental interface or to construct a realistic contact model, but rather to serve as a conceptual tool to inject electrons with a controlled angular-momentum texture, thereby probing the intrinsic filtering response of the nanotube. Within this framework, the coupling is imperfect and is intentionally idealized, and we do not expect---nor require---perfect ballistic transmission. Indeed, none of the injected channels reaches the ideal conductance quantum $G_0 = 2e^2/h$. This suboptimal transmission arises from a structural mismatch between the injected states and the exact crystal-angular-momentum eigenstates of the $(5,5)$ CNT. In our injection scheme, the projected Green's function of Eq.~\eqref{eq:Gl_proj} distributes the phase factors $e^{il\phi_i}$ over $N_\phi = 10$ atomic sites that are \emph{not} uniformly spaced on the circumference (the honeycomb lattice does not tile the circle with equal angular separations), \textit{i.e.}, the orbital texture is discretized using $N_{\phi}=10$ non-uniformly spaced angular sectors. However, the nanotube itself possesses a $C_{5}$ rotational symmetry rather than a $C_{10}$ symmetry, so the discrete Fourier decomposition on 10 irregularly spaced points does not coincide exactly with the irreducible representations of the $C_5$ point group that label the true eigenstates~\cite{BARROS2006261, PhysRevB.93.195442}. Consequently, the injected orbital states exhibit only partial overlap with the CNT's crystal-angular-momentum eigenstates, reducing the mode-matching efficiency and leading to a subsequent reduction in conductance. Importantly, this mismatch does \emph{not} compromise the angular selectivity of the filter: it merely reduces the overall transmission amplitude while preserving the selection rule $l \equiv \mathfrak{m}\pmod{\mathfrak{n}}$. A more refined injection scheme that respects the exact $C_N$ site distribution of the nanotube would yield higher transmission, but the present approach suffices to establish the proof of principle for angular-momentum filtering.

The angular-momentum filtering mechanism discussed above is not restricted to the high-symmetry armchair geometry but extends naturally to other chiral metallic CNT families. To demonstrate this, Fig.~\ref{fig:Filter}(b) presents the conductance of the $(12,3)$ nanotube, a chiral metallic tube of the armchair class with $\mathfrak{n} = \gcd(12,3) = 3$ and, consequently, three distinct angular-momentum sectors $\mathfrak{m} \in \{0, 1, 2\}$. Because this tube belongs to the armchair class, both bands crossing the Fermi level carry $\mathfrak{m} = 0$, and the selection rule \(l \equiv \mathfrak{m} \pmod{\mathfrak{n}}\) dictates that only injected states with $l \equiv 0\pmod{3}$ (\textit{i.e.}, $l = 3, 6, 9, \ldots$) can couple to the low-energy transport channels. Indeed, injection with $l = 3$ yields finite transmission around $E = 0$, while all other tested values ($l = 1, 2, 4, 5$) produce a clear angular-momentum gap at the Fermi level. This result confirms that the filtering operates identically in chiral tubes as in achiral ones: the relevant symmetry is the discrete rotational group $C_{\mathfrak{n}}$, which is common to all CNTs with $\gcd(n,m) = \mathfrak{n}$, rather than any additional mirror or glide symmetry specific to armchair or zigzag geometries~\cite{PhysRevB.93.195442, BARROS2006261}. Notably, the lower value of $\mathfrak{n} = 3$ compared to $\mathfrak{n} = 5$ for the $(5,5)$ tube means that fewer angular-momentum channels are available, resulting in a coarser filtering spectrum but also in wider angular-momentum gaps, a trade-off between spectral resolution and gap robustness that may be exploited in device design.

Finally, we consider the zigzag-class metallic CNT $(12,6)$ [$\mathfrak{n} = \gcd(12,6) = 6$], shown in Fig.~\ref{fig:Filter}(c). Unlike the armchair-class examples discussed above, the zigzag class features two degenerate low-energy bands with \emph{distinct} crystal angular momenta: $\mathfrak{m}_a = (2\times12 + 6)/3\;\mathrm{mod}\;6 = 4$ and $\mathfrak{m}_b = (2\times6 + 12)/3\;\mathrm{mod}\;6 = 2$~\cite{Lunde}. The selection rule, therefore, admits two allowed injection channels near the Fermi level: $l \equiv 4\pmod{6}$ and $l \equiv 2\pmod{6}$. Indeed, the calculated conductance confirms that only $l = 4$ and $l = 2$ yield finite transmission at low energies, while all other injected angular momenta are completely blocked. This result demonstrates a qualitatively richer filtering landscape than the armchair-class case: because the two Fermi-level channels carry different $\mathfrak{m}$ values, the nanotube can \emph{selectively} transmit one channel while blocking the other, depending on the angular momentum of the injected state. In other words, the zigzag-class chiral CNT operates as a true angular-momentum beam splitter, injecting $l = 4$ addresses exclusively to the $\mathfrak{m}_a$ channel, while $l = 2$ addresses only the $\mathfrak{m}_b$, with no cross-talk between them~\cite{PhysRevB.93.195442, PhysRevB.92.075433}. This channel-selective transparency, governed solely by the discrete rotational symmetry and the chiral indices, establishes chiral CNTs as minimal components for orbital-angular-momentum demultiplexing at the nanoscale.

\section{Final Remarks}

In summary, we have presented a theoretical study of the 
non-equilibrium orbital Edelstein effect in chiral carbon nanotubes, encompassing 86 distinct chiralities across both metallic and semiconducting families. From a coarse-grained perspective, our results demonstrate that CNTs exhibit a non-equilibrium chirality-dependent orbital angular momentum selectivity, whereby charge currents generate orbital magnetization whose sign and magnitude are controlled by the nanotube chirality, governed by the chiral wrapping vector, being a manifestation of CIOS in a realistic quasi-one-dimensional material~\cite{Gobel_CIOS, BahamonChV3}. A more detailed analysis, however, reveals a much richer behavior that goes beyond the equilibrium picture commonly inferred from the graphene band structure \cite{Gobel_CNT, PhysRevB.78.233405, PhysRevLett.76.2121}. Three principal findings emerge. First, we showed that although the orbital Edelstein susceptibility generally increases with nanotube diameter \cite{McEuen_muexp}, the data do not collapse onto a universal scaling curve, \textit{i.e.}, $\chi_z^{L_z}$ does not obey a universal diameter scaling, but rather the data segregate into distinct family-dependent branches, analogous to the patterns observed in optical Kataura plots~\cite{KATAURA19992555, Bachilo2002}, demonstrating that nanotubes of nearly identical diameter but different chiral indices can exhibit orbital responses differing by up to a factor of two. This branch structure reflects the subband-resolved angular-momentum texture unique to each $(n,m)$ chirality and establishes the orbital susceptibility as a sensitive spectroscopic fingerprint of the nanotube's electronic identity. Second, the role of metallic contacts differs qualitatively between metallic and semiconducting CNTs. In metallic tubes, wide-band leads inject angularly unpolarized current that is rapidly projected onto the nanotube's intrinsic angular-momentum eigenstates within three to four unit cells ($\sim$3--5~nm), recovering the bulk orbital response with remarkable efficiency. In semiconducting CNT, by contrast, the coherent excitation of multiple channels with distinct crystal angular momenta gives rise to persistent spatial oscillations, a quantum beating phenomenon that can generate transverse orbital moments exceeding the longitudinal Edelstein response. These observations highlight that contact engineering is not merely a practical necessity but an active degree of freedom for shaping the orbital magnetization texture in CNT-based orbitronic devices. Third, by introducing angular correlations into the contact self-energy, we have demonstrated that chiral CNTs operate as efficient orbital-angular-momentum filters governed by the selection rule $l \equiv \mathfrak{m}\pmod{\mathfrak{n}}$. The filtering mechanism relies solely on the discrete rotational symmetry $C_{\mathfrak{n}}$ of the nanotube and applies equally to achiral and genuinely chiral tubes, establishing CNTs as minimal nanoscale components for angular-momentum demultiplexing. Taken together, our findings demonstrate that the non-equilibrium orbital properties of CNTs are governed not only by their diameter and chirality, but also by the detailed electronic structure and contact geometry of the device.

\section*{Acknowledgments}

S.S. gratefully acknowledges CAPES-PROSUC (88887.192769/2025-00). D.A.B. acknowledges support from INCT de Nanomateriais de Carbono e Materiais 2D (INCT/Nanocarbon), CNPq (Grant No. 304626/2024-4), FAPESP (Grant Nos. 23/13081-0 and 25/17579-9), and MackPesquisa. D.~R.~C. gratefully acknowledges the Brazilian National Council for the Improvement of Higher Education (CAPES), the support from CNPq Grant No. $312539/2025$--$8$, No. $437067/2018$--$1$, No. $423423/2021$--$5$, No. $408144/2022$--$0$, and the Fundação Cearense de Apoio ao Desenvolvimento Científico e Tecnológico (FUNCAP).

\section*{DATA AVAILABILITY}

The data that support the findings of this study are available from the corresponding author upon reasonable request.
\appendix
\section{Effect of Curvature}
\label{sec.appendix}


\begin{center}
\begin{figure}[!t]
\includegraphics[width=\linewidth]{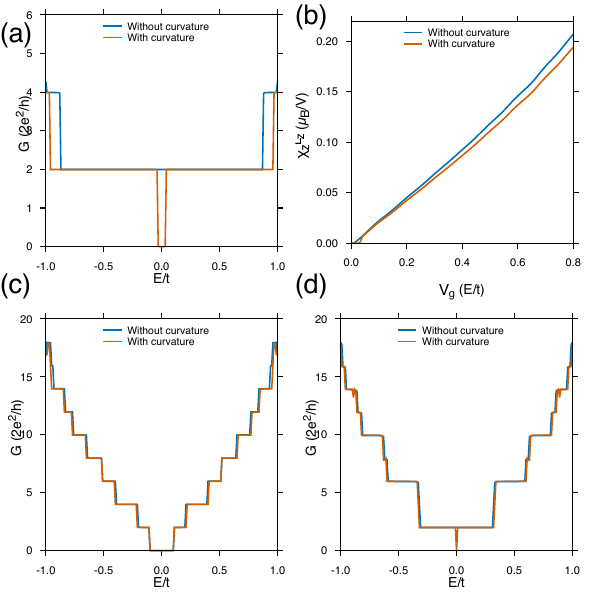}
\caption{Comparison of results obtained without curvature (blue) and with curvature included via the Slater--Koster parametrization (orange). (a)~Two-terminal conductance of the metallic $(4,1)$ CNT ($d = 3.58$~\AA), showing a curvature-induced gap of approximately $0.03\,t$ around the charge-neutrality point; away from this narrow region, the transmission is essentially unaffected. (b)~Orbital Edelstein susceptibility $\chi_z^{L_z}$ of the same $(4,1)$ CNT as a function of gate voltage: the overall lineshape and monotonic increase are preserved, with only a modest quantitative reduction due to the suppression of transport within the curvature-induced gap. (c)~Conductance of the semiconducting $(12,8)$ CNT ($d = 13.6$~\AA) and (d)~conductance of the metallic $(11,8)$ CNT ($d = 12.9$~\AA).}
\label{fig:curvature} 
\end{figure}
\end{center}

\begin{center}
\begin{figure}[!t]
\scalebox{0.85}{\includegraphics{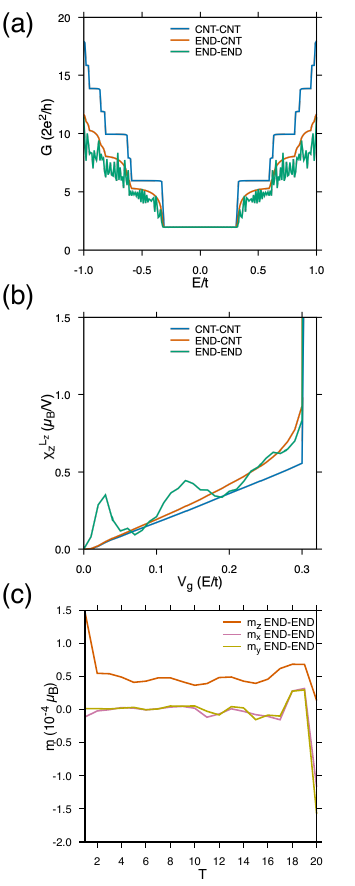}}
\caption{Effect of introducing a second metallic contact on the right side,
        with both contacts in the END configuration, for the metallic
        $(11,8)$ CNT. (a) Conductance $G$ as a function of the Fermi energy
        $E/t$ for the CNT--CNT, END--CNT, and END--END contact configurations.
        (b) Edelstein susceptibility $\chi_z^{L}$ as a function of the gate
        voltage $V_g$ for the same three contact configurations.
        (c) Spatial dependence of the components of the average magnetic
        angular momentum per unit cell, $m_x$, $m_y$, and $m_z$, along a CNT
        consisting of 20 unit cells for the END--END configuration ($V_g = 0.25t$). The
        metallic contacts at both ends give rise to back-propagating waves
        and interference effects, which produce oscillations in the transport
        and magnetic response.}
\label{fig:fig7} 
\end{figure}
\end{center}

To quantify the impact of curvature on our results, we performed 
additional calculations incorporating its  contribution within 
the single-orbital framework through a Slater--Koster parametrization 
of the hopping integrals~\cite{SlaterKoster,KleinerEggert2001}:

\begin{equation}
t_{ij} = V_{pp\pi}\,\hat{\mathbf{n}}_i\cdot\hat{\mathbf{n}}_j 
+ (V_{pp\sigma} - V_{pp\pi})
(\hat{\mathbf{n}}_i\cdot\hat{\mathbf{d}}_{ij})
(\hat{\mathbf{n}}_j\cdot\hat{\mathbf{d}}_{ij}),
\end{equation}

where $\hat{\mathbf{n}}_i$ and $\hat{\mathbf{n}}_j$ are the local radial 
unit vectors at sites $i$ and $j$, respectively, specifying the 
orientation of the corresponding $p_z$ orbitals, and 
$\hat{\mathbf{d}}_{ij}$ is the unit vector along the bond 
connecting the two neighboring carbon atoms. We use 
$V_{pp\pi}=-2.77$~eV and $V_{pp\sigma}=4.37$~eV~\cite{WhiteMintmire1998}.

As shown in Fig.~\ref{fig:curvature}, curvature produces an appreciable 
modification only for the smallest metallic CNT in our dataset, the 
$(4,1)$ nanotube ($d=3.58$~\AA), where it opens a half-gap of 
approximately $0.03t$. Nevertheless, the orbital Edelstein susceptibility 
retains its overall lineshape and monotonic increase with $V_g$, with 
the main change being a suppression of the response within the 
curvature-induced gap. For larger-diameter metallic CNTs, such as the 
$(11,8)$ nanotube, the much smaller gap has no appreciable influence on 
the current-induced response after the energy integration. The same 
behavior is found for the semiconducting $(12,8)$ CNT. Thus, including 
curvature does not modify the chirality-dependent trends reported in 
this work.

We note that a full treatment of $\sigma$--$\pi$ rehybridization 
requires the explicit inclusion of $\sigma$ orbitals and is particularly 
relevant for optical properties~\cite{bookSaito,Reich2004}. For the 
low-energy transport regime considered here, the $p_z$-only model 
supplemented by the Slater--Koster curvature correction provides an 
adequate description.

\section{Effect of a Second Metallic Contact at the Right End}
\label{sec.appB}

In the main text, we explain that we deliberately inject charge through a 
single metallic contact on the left in order to minimize interference effects 
within the CNT, allowing the angular-momentum selectivity to be examined more 
clearly. In this section, for completeness, we consider the metallic $(11,8)$ 
CNT with metallic contacts at both the left and right ends in the end-contact 
configuration, and present the corresponding conductance, Edelstein 
susceptibility, and average magnetic angular momentum. As expected, the 
additional metallic contact on the right generates back-propagating waves, 
leading to characteristic oscillatory features. These effects can be directly 
contrasted with the results shown in Figs.~\ref{fig:endside}(a), (c), and (e), 
which correspond to the configuration with a single metallic contact on the 
left.

Specifically, these oscillatory features are observed in the conductance 
[Fig.~\ref{fig:fig7}(a)], the Edelstein susceptibility 
[Fig.~\ref{fig:fig7}(b)], and the average magnetic angular momentum per unit 
cell [Fig.~\ref{fig:fig7}(c)]. In the conductance, although the selected scale 
gives the appearance of an approximately flat plateau, a closer inspection 
reveals oscillations resulting from the superposition of forward- and 
backward-propagating waves. 

The effect of the second metallic contact is also evident in the spatial 
distribution of the average orbital magnetic moment. For the $20$-unit-cell 
CNT, the $m_x$ and $m_y$ components are enhanced near the entrance and exit 
of the CNT, while all three components, $m_x$, $m_y$, and $m_z$, exhibit 
 spatial oscillations throughout the device. This behavior contrasts 
with the single-metal-contact configuration discussed in the main text, for 
which the $(11,8)$ CNT reaches a smooth spatial profile approximately four 
unit cells away from the metallic contact. The persistence of the oscillations 
throughout the CNT in the two-metal-contact configuration is consistent with 
the interference between forward- and backward-propagating electronic states 
generated by the second metallic contact. This result corroborates our initial 
rationale for employing a single metallic contact: by suppressing these 
additional oscillatory contributions, the single-contact configuration 
provides a cleaner setting for analyzing the angular-momentum selectivity of 
the CNT.

\bibliography{references_resubmitted.bib}

\begin{thebibliography}{74}%
\makeatletter
\providecommand \@ifxundefined [1]{%
 \@ifx{#1\undefined}
}%
\providecommand \@ifnum [1]{%
 \ifnum #1\expandafter \@firstoftwo
 \else \expandafter \@secondoftwo
 \fi
}%
\providecommand \@ifx [1]{%
 \ifx #1\expandafter \@firstoftwo
 \else \expandafter \@secondoftwo
 \fi
}%
\providecommand \natexlab [1]{#1}%
\providecommand \enquote  [1]{``#1''}%
\providecommand \bibnamefont  [1]{#1}%
\providecommand \bibfnamefont [1]{#1}%
\providecommand \citenamefont [1]{#1}%
\providecommand \href@noop [0]{\@secondoftwo}%
\providecommand \href [0]{\begingroup \@sanitize@url \@href}%
\providecommand \@href[1]{\@@startlink{#1}\@@href}%
\providecommand \@@href[1]{\endgroup#1\@@endlink}%
\providecommand \@sanitize@url [0]{\catcode `\\12\catcode `\$12\catcode
  `\&12\catcode `\#12\catcode `\^12\catcode `\_12\catcode `\%12\relax}%
\providecommand \@@startlink[1]{}%
\providecommand \@@endlink[0]{}%
\providecommand \url  [0]{\begingroup\@sanitize@url \@url }%
\providecommand \@url [1]{\endgroup\@href {#1}{\urlprefix }}%
\providecommand \urlprefix  [0]{URL }%
\providecommand \Eprint [0]{\href }%
\providecommand \doibase [0]{https://doi.org/}%
\providecommand \selectlanguage [0]{\@gobble}%
\providecommand \bibinfo  [0]{\@secondoftwo}%
\providecommand \bibfield  [0]{\@secondoftwo}%
\providecommand \translation [1]{[#1]}%
\providecommand \BibitemOpen [0]{}%
\providecommand \bibitemStop [0]{}%
\providecommand \bibitemNoStop [0]{.\EOS\space}%
\providecommand \EOS [0]{\spacefactor3000\relax}%
\providecommand \BibitemShut  [1]{\csname bibitem#1\endcsname}%
\let\auto@bib@innerbib\@empty
\bibitem [{\citenamefont {Yan}(2024)}]{Binghai_review}%
  \BibitemOpen
  \bibfield  {author} {\bibinfo {author} {\bibfnamefont {B.}~\bibnamefont
  {Yan}},\ }\bibfield  {title} {\bibinfo {title} {Structural chirality and
  electronic chirality in quantum materials},\ }\href
  {https://doi.org/https://doi.org/10.1146/annurev-matsci-080222-033548}
  {\bibfield  {journal} {\bibinfo  {journal} {Annual Review of Materials
  Research}\ }\textbf {\bibinfo {volume} {54}},\ \bibinfo {pages} {97}
  (\bibinfo {year} {2024})}\BibitemShut {NoStop}%
\bibitem [{\citenamefont {Aiello}\ \emph {et~al.}(2022)\citenamefont {Aiello},
  \citenamefont {Abendroth}, \citenamefont {Abbas}, \citenamefont {Afanasev},
  \citenamefont {Agarwal}, \citenamefont {Banerjee}, \citenamefont {Beratan},
  \citenamefont {Belling}, \citenamefont {Berche}, \citenamefont {Botana},
  \citenamefont {Caram}, \citenamefont {Celardo}, \citenamefont {Cuniberti},
  \citenamefont {Garcia-Etxarri}, \citenamefont {Dianat}, \citenamefont
  {Diez-Perez}, \citenamefont {Guo}, \citenamefont {Gutierrez}, \citenamefont
  {Herrmann}, \citenamefont {Hihath}, \citenamefont {Kale}, \citenamefont
  {Kurian}, \citenamefont {Lai}, \citenamefont {Liu}, \citenamefont {Lopez},
  \citenamefont {Medina}, \citenamefont {Mujica}, \citenamefont {Naaman},
  \citenamefont {Noormandipour}, \citenamefont {Palma}, \citenamefont
  {Paltiel}, \citenamefont {Petuskey}, \citenamefont {Ribeiro-Silva},
  \citenamefont {Saenz}, \citenamefont {Santos}, \citenamefont
  {Solyanik-Gorgone}, \citenamefont {Sorger}, \citenamefont {Stemer},
  \citenamefont {Ugalde}, \citenamefont {Valdes-Curiel}, \citenamefont
  {Varela}, \citenamefont {Waldeck}, \citenamefont {Wasielewski}, \citenamefont
  {Weiss}, \citenamefont {Zacharias},\ and\ \citenamefont
  {Wang}}]{Chirality_quantumLeap}%
  \BibitemOpen
  \bibfield  {author} {\bibinfo {author} {\bibfnamefont {C.~D.}\ \bibnamefont
  {Aiello}}, \bibinfo {author} {\bibfnamefont {J.~M.}\ \bibnamefont
  {Abendroth}}, \bibinfo {author} {\bibfnamefont {M.}~\bibnamefont {Abbas}},
  \bibinfo {author} {\bibfnamefont {A.}~\bibnamefont {Afanasev}}, \bibinfo
  {author} {\bibfnamefont {S.}~\bibnamefont {Agarwal}}, \bibinfo {author}
  {\bibfnamefont {A.~S.}\ \bibnamefont {Banerjee}}, \bibinfo {author}
  {\bibfnamefont {D.~N.}\ \bibnamefont {Beratan}}, \bibinfo {author}
  {\bibfnamefont {J.~N.}\ \bibnamefont {Belling}}, \bibinfo {author}
  {\bibfnamefont {B.}~\bibnamefont {Berche}}, \bibinfo {author} {\bibfnamefont
  {A.}~\bibnamefont {Botana}}, \bibinfo {author} {\bibfnamefont {J.~R.}\
  \bibnamefont {Caram}}, \bibinfo {author} {\bibfnamefont {G.~L.}\ \bibnamefont
  {Celardo}}, \bibinfo {author} {\bibfnamefont {G.}~\bibnamefont {Cuniberti}},
  \bibinfo {author} {\bibfnamefont {A.}~\bibnamefont {Garcia-Etxarri}},
  \bibinfo {author} {\bibfnamefont {A.}~\bibnamefont {Dianat}}, \bibinfo
  {author} {\bibfnamefont {I.}~\bibnamefont {Diez-Perez}}, \bibinfo {author}
  {\bibfnamefont {Y.}~\bibnamefont {Guo}}, \bibinfo {author} {\bibfnamefont
  {R.}~\bibnamefont {Gutierrez}}, \bibinfo {author} {\bibfnamefont
  {C.}~\bibnamefont {Herrmann}}, \bibinfo {author} {\bibfnamefont
  {J.}~\bibnamefont {Hihath}}, \bibinfo {author} {\bibfnamefont
  {S.}~\bibnamefont {Kale}}, \bibinfo {author} {\bibfnamefont {P.}~\bibnamefont
  {Kurian}}, \bibinfo {author} {\bibfnamefont {Y.-C.}\ \bibnamefont {Lai}},
  \bibinfo {author} {\bibfnamefont {T.}~\bibnamefont {Liu}}, \bibinfo {author}
  {\bibfnamefont {A.}~\bibnamefont {Lopez}}, \bibinfo {author} {\bibfnamefont
  {E.}~\bibnamefont {Medina}}, \bibinfo {author} {\bibfnamefont
  {V.}~\bibnamefont {Mujica}}, \bibinfo {author} {\bibfnamefont
  {R.}~\bibnamefont {Naaman}}, \bibinfo {author} {\bibfnamefont
  {M.}~\bibnamefont {Noormandipour}}, \bibinfo {author} {\bibfnamefont {J.~L.}\
  \bibnamefont {Palma}}, \bibinfo {author} {\bibfnamefont {Y.}~\bibnamefont
  {Paltiel}}, \bibinfo {author} {\bibfnamefont {W.}~\bibnamefont {Petuskey}},
  \bibinfo {author} {\bibfnamefont {J.~C.}\ \bibnamefont {Ribeiro-Silva}},
  \bibinfo {author} {\bibfnamefont {J.~J.}\ \bibnamefont {Saenz}}, \bibinfo
  {author} {\bibfnamefont {E.~J.~G.}\ \bibnamefont {Santos}}, \bibinfo {author}
  {\bibfnamefont {M.}~\bibnamefont {Solyanik-Gorgone}}, \bibinfo {author}
  {\bibfnamefont {V.~J.}\ \bibnamefont {Sorger}}, \bibinfo {author}
  {\bibfnamefont {D.~M.}\ \bibnamefont {Stemer}}, \bibinfo {author}
  {\bibfnamefont {J.~M.}\ \bibnamefont {Ugalde}}, \bibinfo {author}
  {\bibfnamefont {A.}~\bibnamefont {Valdes-Curiel}}, \bibinfo {author}
  {\bibfnamefont {S.}~\bibnamefont {Varela}}, \bibinfo {author} {\bibfnamefont
  {D.~H.}\ \bibnamefont {Waldeck}}, \bibinfo {author} {\bibfnamefont {M.~R.}\
  \bibnamefont {Wasielewski}}, \bibinfo {author} {\bibfnamefont {P.~S.}\
  \bibnamefont {Weiss}}, \bibinfo {author} {\bibfnamefont {H.}~\bibnamefont
  {Zacharias}},\ and\ \bibinfo {author} {\bibfnamefont {Q.~H.}\ \bibnamefont
  {Wang}},\ }\bibfield  {title} {\bibinfo {title} {A chirality-based quantum
  leap},\ }\href {https://doi.org/10.1021/acsnano.1c01347} {\bibfield
  {journal} {\bibinfo  {journal} {ACS Nano}\ }\textbf {\bibinfo {volume}
  {16}},\ \bibinfo {pages} {4989} (\bibinfo {year} {2022})}\BibitemShut
  {NoStop}%
\bibitem [{\citenamefont {Bloom}\ \emph {et~al.}(2024)\citenamefont {Bloom},
  \citenamefont {Paltiel}, \citenamefont {Naaman},\ and\ \citenamefont
  {Waldeck}}]{Review_CISS}%
  \BibitemOpen
  \bibfield  {author} {\bibinfo {author} {\bibfnamefont {B.~P.}\ \bibnamefont
  {Bloom}}, \bibinfo {author} {\bibfnamefont {Y.}~\bibnamefont {Paltiel}},
  \bibinfo {author} {\bibfnamefont {R.}~\bibnamefont {Naaman}},\ and\ \bibinfo
  {author} {\bibfnamefont {D.~H.}\ \bibnamefont {Waldeck}},\ }\bibfield
  {title} {\bibinfo {title} {Chiral induced spin selectivity},\ }\href
  {https://doi.org/10.1021/acs.chemrev.3c00661} {\bibfield  {journal} {\bibinfo
   {journal} {Chemical Reviews}\ }\textbf {\bibinfo {volume} {124}},\ \bibinfo
  {pages} {1950} (\bibinfo {year} {2024})}\BibitemShut {NoStop}%
\bibitem [{\citenamefont {Tokura}\ and\ \citenamefont
  {Nagaosa}(2018)}]{NRTokura}%
  \BibitemOpen
  \bibfield  {author} {\bibinfo {author} {\bibfnamefont {Y.}~\bibnamefont
  {Tokura}}\ and\ \bibinfo {author} {\bibfnamefont {N.}~\bibnamefont
  {Nagaosa}},\ }\bibfield  {title} {\bibinfo {title} {Nonreciprocal responses
  from non-centrosymmetric quantum materials},\ }\href
  {https://doi.org/10.1038/s41467-018-05759-4} {\bibfield  {journal} {\bibinfo
  {journal} {Nature Communications}\ }\textbf {\bibinfo {volume} {9}},\
  \bibinfo {pages} {3740} (\bibinfo {year} {2018})}\BibitemShut {NoStop}%
\bibitem [{\citenamefont {Rikken}\ and\ \citenamefont
  {Avarvari}(2023)}]{doi:10.1021/acs.jpclett.3c02546}%
  \BibitemOpen
  \bibfield  {author} {\bibinfo {author} {\bibfnamefont {G.~L. J.~A.}\
  \bibnamefont {Rikken}}\ and\ \bibinfo {author} {\bibfnamefont
  {N.}~\bibnamefont {Avarvari}},\ }\bibfield  {title} {\bibinfo {title}
  {Comparing electrical magnetochiral anisotropy and chirality-induced spin
  selectivity},\ }\href {https://doi.org/10.1021/acs.jpclett.3c02546}
  {\bibfield  {journal} {\bibinfo  {journal} {The Journal of Physical Chemistry
  Letters}\ }\textbf {\bibinfo {volume} {14}},\ \bibinfo {pages} {9727}
  (\bibinfo {year} {2023})}\BibitemShut {NoStop}%
\bibitem [{\citenamefont {Naaman}\ \emph {et~al.}(2019)\citenamefont {Naaman},
  \citenamefont {Paltiel},\ and\ \citenamefont {Waldeck}}]{Naaman2019}%
  \BibitemOpen
  \bibfield  {author} {\bibinfo {author} {\bibfnamefont {R.}~\bibnamefont
  {Naaman}}, \bibinfo {author} {\bibfnamefont {Y.}~\bibnamefont {Paltiel}},\
  and\ \bibinfo {author} {\bibfnamefont {D.~H.}\ \bibnamefont {Waldeck}},\
  }\bibfield  {title} {\bibinfo {title} {Chiral molecules and the electron
  spin},\ }\href {https://doi.org/10.1038/s41570-019-0087-1} {\bibfield
  {journal} {\bibinfo  {journal} {Nature Reviews Chemistry}\ }\textbf {\bibinfo
  {volume} {3}},\ \bibinfo {pages} {250} (\bibinfo {year} {2019})}\BibitemShut
  {NoStop}%
\bibitem [{\citenamefont {Evers}\ \emph {et~al.}(2022)\citenamefont {Evers},
  \citenamefont {Aharony}, \citenamefont {Bar-Gill}, \citenamefont
  {Entin-Wohlman}, \citenamefont {Hedegård}, \citenamefont {Hod},
  \citenamefont {Jelinek}, \citenamefont {Kamieniarz}, \citenamefont
  {Lemeshko}, \citenamefont {Michaeli}, \citenamefont {Mujica}, \citenamefont
  {Naaman}, \citenamefont {Paltiel}, \citenamefont {Refaely-Abramson},
  \citenamefont {Tal}, \citenamefont {Thijssen}, \citenamefont {Thoss},
  \citenamefont {van Ruitenbeek}, \citenamefont {Venkataraman}, \citenamefont
  {Waldeck}, \citenamefont {Yan},\ and\ \citenamefont {Kronik}}]{Evers2022}%
  \BibitemOpen
  \bibfield  {author} {\bibinfo {author} {\bibfnamefont {F.}~\bibnamefont
  {Evers}}, \bibinfo {author} {\bibfnamefont {A.}~\bibnamefont {Aharony}},
  \bibinfo {author} {\bibfnamefont {N.}~\bibnamefont {Bar-Gill}}, \bibinfo
  {author} {\bibfnamefont {O.}~\bibnamefont {Entin-Wohlman}}, \bibinfo {author}
  {\bibfnamefont {P.}~\bibnamefont {Hedegård}}, \bibinfo {author}
  {\bibfnamefont {O.}~\bibnamefont {Hod}}, \bibinfo {author} {\bibfnamefont
  {P.}~\bibnamefont {Jelinek}}, \bibinfo {author} {\bibfnamefont
  {G.}~\bibnamefont {Kamieniarz}}, \bibinfo {author} {\bibfnamefont
  {M.}~\bibnamefont {Lemeshko}}, \bibinfo {author} {\bibfnamefont
  {K.}~\bibnamefont {Michaeli}}, \bibinfo {author} {\bibfnamefont
  {V.}~\bibnamefont {Mujica}}, \bibinfo {author} {\bibfnamefont
  {R.}~\bibnamefont {Naaman}}, \bibinfo {author} {\bibfnamefont
  {Y.}~\bibnamefont {Paltiel}}, \bibinfo {author} {\bibfnamefont
  {S.}~\bibnamefont {Refaely-Abramson}}, \bibinfo {author} {\bibfnamefont
  {O.}~\bibnamefont {Tal}}, \bibinfo {author} {\bibfnamefont {J.}~\bibnamefont
  {Thijssen}}, \bibinfo {author} {\bibfnamefont {M.}~\bibnamefont {Thoss}},
  \bibinfo {author} {\bibfnamefont {J.~M.}\ \bibnamefont {van Ruitenbeek}},
  \bibinfo {author} {\bibfnamefont {L.}~\bibnamefont {Venkataraman}}, \bibinfo
  {author} {\bibfnamefont {D.~H.}\ \bibnamefont {Waldeck}}, \bibinfo {author}
  {\bibfnamefont {B.}~\bibnamefont {Yan}},\ and\ \bibinfo {author}
  {\bibfnamefont {L.}~\bibnamefont {Kronik}},\ }\bibfield  {title} {\bibinfo
  {title} {Theory of chirality induced spin selectivity: Progress and
  challenges},\ }\href {https://doi.org/https://doi.org/10.1002/adma.202106629}
  {\bibfield  {journal} {\bibinfo  {journal} {Advanced Materials}\ }\textbf
  {\bibinfo {volume} {34}},\ \bibinfo {pages} {2106629} (\bibinfo {year}
  {2022})}\BibitemShut {NoStop}%
\bibitem [{\citenamefont {Liu}\ \emph {et~al.}(2021{\natexlab{a}})\citenamefont
  {Liu}, \citenamefont {Xiao}, \citenamefont {Koo},\ and\ \citenamefont
  {Yan}}]{Liu2021spinfilter}%
  \BibitemOpen
  \bibfield  {author} {\bibinfo {author} {\bibfnamefont {Y.}~\bibnamefont
  {Liu}}, \bibinfo {author} {\bibfnamefont {J.}~\bibnamefont {Xiao}}, \bibinfo
  {author} {\bibfnamefont {J.}~\bibnamefont {Koo}},\ and\ \bibinfo {author}
  {\bibfnamefont {B.}~\bibnamefont {Yan}},\ }\bibfield  {title} {\bibinfo
  {title} {Chirality-driven topological electronic structure of {DNA}-like
  materials},\ }\href {https://doi.org/10.1038/s41563-021-00924-5} {\bibfield
  {journal} {\bibinfo  {journal} {Nature Materials}\ }\textbf {\bibinfo
  {volume} {20}},\ \bibinfo {pages} {638} (\bibinfo {year}
  {2021}{\natexlab{a}})}\BibitemShut {NoStop}%
\bibitem [{\citenamefont {Rikken}\ \emph {et~al.}(2001)\citenamefont {Rikken},
  \citenamefont {F\"olling},\ and\ \citenamefont {Wyder}}]{Rikken2001}%
  \BibitemOpen
  \bibfield  {author} {\bibinfo {author} {\bibfnamefont {G.~L. J.~A.}\
  \bibnamefont {Rikken}}, \bibinfo {author} {\bibfnamefont {J.}~\bibnamefont
  {F\"olling}},\ and\ \bibinfo {author} {\bibfnamefont {P.}~\bibnamefont
  {Wyder}},\ }\bibfield  {title} {\bibinfo {title} {Electrical magnetochiral
  anisotropy},\ }\href {https://doi.org/10.1103/PhysRevLett.87.236602}
  {\bibfield  {journal} {\bibinfo  {journal} {Physical Review Letters}\
  }\textbf {\bibinfo {volume} {87}},\ \bibinfo {pages} {236602} (\bibinfo
  {year} {2001})}\BibitemShut {NoStop}%
\bibitem [{\citenamefont {Pop}\ \emph {et~al.}(2014)\citenamefont {Pop},
  \citenamefont {Auban-Senzier}, \citenamefont {Canadell}, \citenamefont
  {Rikken},\ and\ \citenamefont {Avarvari}}]{Pop2014}%
  \BibitemOpen
  \bibfield  {author} {\bibinfo {author} {\bibfnamefont {F.}~\bibnamefont
  {Pop}}, \bibinfo {author} {\bibfnamefont {P.}~\bibnamefont {Auban-Senzier}},
  \bibinfo {author} {\bibfnamefont {E.}~\bibnamefont {Canadell}}, \bibinfo
  {author} {\bibfnamefont {G.~L. J.~A.}\ \bibnamefont {Rikken}},\ and\ \bibinfo
  {author} {\bibfnamefont {N.}~\bibnamefont {Avarvari}},\ }\bibfield  {title}
  {\bibinfo {title} {Electrical magnetochiral anisotropy in a bulk chiral
  molecular conductor},\ }\href {https://doi.org/10.1038/ncomms4757} {\bibfield
   {journal} {\bibinfo  {journal} {Nature Communications}\ }\textbf {\bibinfo
  {volume} {5}},\ \bibinfo {pages} {3757} (\bibinfo {year} {2014})}\BibitemShut
  {NoStop}%
\bibitem [{\citenamefont {Yokouchi}\ \emph {et~al.}(2017)\citenamefont
  {Yokouchi}, \citenamefont {Kanazawa}, \citenamefont {Kikkawa}, \citenamefont
  {Morikawa}, \citenamefont {Shibata}, \citenamefont {Arima}, \citenamefont
  {Taguchi}, \citenamefont {Kagawa},\ and\ \citenamefont
  {Tokura}}]{Yokouchi2017}%
  \BibitemOpen
  \bibfield  {author} {\bibinfo {author} {\bibfnamefont {T.}~\bibnamefont
  {Yokouchi}}, \bibinfo {author} {\bibfnamefont {N.}~\bibnamefont {Kanazawa}},
  \bibinfo {author} {\bibfnamefont {A.}~\bibnamefont {Kikkawa}}, \bibinfo
  {author} {\bibfnamefont {D.}~\bibnamefont {Morikawa}}, \bibinfo {author}
  {\bibfnamefont {K.}~\bibnamefont {Shibata}}, \bibinfo {author} {\bibfnamefont
  {T.-h.}\ \bibnamefont {Arima}}, \bibinfo {author} {\bibfnamefont
  {Y.}~\bibnamefont {Taguchi}}, \bibinfo {author} {\bibfnamefont
  {F.}~\bibnamefont {Kagawa}},\ and\ \bibinfo {author} {\bibfnamefont
  {Y.}~\bibnamefont {Tokura}},\ }\bibfield  {title} {\bibinfo {title}
  {Electrical magnetochiral effect induced by chiral spin fluctuations},\
  }\href {https://doi.org/10.1038/s41467-017-01094-2} {\bibfield  {journal}
  {\bibinfo  {journal} {Nature Communications}\ }\textbf {\bibinfo {volume}
  {8}},\ \bibinfo {pages} {866} (\bibinfo {year} {2017})}\BibitemShut {NoStop}%
\bibitem [{\citenamefont {Ideue}\ \emph {et~al.}(2017)\citenamefont {Ideue},
  \citenamefont {Hamamoto}, \citenamefont {Koshikawa}, \citenamefont {Ezawa},
  \citenamefont {Shimizu}, \citenamefont {Kaneko}, \citenamefont {Tokura},
  \citenamefont {Nagaosa},\ and\ \citenamefont {Iwasa}}]{Ideue2017}%
  \BibitemOpen
  \bibfield  {author} {\bibinfo {author} {\bibfnamefont {T.}~\bibnamefont
  {Ideue}}, \bibinfo {author} {\bibfnamefont {K.}~\bibnamefont {Hamamoto}},
  \bibinfo {author} {\bibfnamefont {S.}~\bibnamefont {Koshikawa}}, \bibinfo
  {author} {\bibfnamefont {M.}~\bibnamefont {Ezawa}}, \bibinfo {author}
  {\bibfnamefont {S.}~\bibnamefont {Shimizu}}, \bibinfo {author} {\bibfnamefont
  {Y.}~\bibnamefont {Kaneko}}, \bibinfo {author} {\bibfnamefont
  {Y.}~\bibnamefont {Tokura}}, \bibinfo {author} {\bibfnamefont
  {N.}~\bibnamefont {Nagaosa}},\ and\ \bibinfo {author} {\bibfnamefont
  {Y.}~\bibnamefont {Iwasa}},\ }\bibfield  {title} {\bibinfo {title} {Bulk
  rectification effect in a polar semiconductor},\ }\href
  {https://doi.org/10.1038/nphys4056} {\bibfield  {journal} {\bibinfo
  {journal} {Nature Physics}\ }\textbf {\bibinfo {volume} {13}},\ \bibinfo
  {pages} {578} (\bibinfo {year} {2017})}\BibitemShut {NoStop}%
\bibitem [{\citenamefont {Bahamon}\ \emph {et~al.}(2024)\citenamefont
  {Bahamon}, \citenamefont {Gómez-Santos}, \citenamefont {Efetov},\ and\
  \citenamefont {Stauber}}]{BahamonChV3}%
  \BibitemOpen
  \bibfield  {author} {\bibinfo {author} {\bibfnamefont {D.~A.}\ \bibnamefont
  {Bahamon}}, \bibinfo {author} {\bibfnamefont {G.}~\bibnamefont
  {Gómez-Santos}}, \bibinfo {author} {\bibfnamefont {D.~K.}\ \bibnamefont
  {Efetov}},\ and\ \bibinfo {author} {\bibfnamefont {T.}~\bibnamefont
  {Stauber}},\ }\bibfield  {title} {\bibinfo {title} {Chirality probe of
  twisted bilayer graphene in the linear transport regime},\ }\href
  {https://doi.org/10.1021/acs.nanolett.4c00371} {\bibfield  {journal}
  {\bibinfo  {journal} {Nano Letters}\ }\textbf {\bibinfo {volume} {24}},\
  \bibinfo {pages} {4478} (\bibinfo {year} {2024})}\BibitemShut {NoStop}%
\bibitem [{\citenamefont {Kuhn}\ \emph {et~al.}(2025)\citenamefont {Kuhn},
  \citenamefont {Silva},\ and\ \citenamefont {Bahamon}}]{MatheusCount}%
  \BibitemOpen
  \bibfield  {author} {\bibinfo {author} {\bibfnamefont {M.~H.~G.}\
  \bibnamefont {Kuhn}}, \bibinfo {author} {\bibfnamefont {L.~A.}\ \bibnamefont
  {Silva}},\ and\ \bibinfo {author} {\bibfnamefont {D.~A.}\ \bibnamefont
  {Bahamon}},\ }\bibfield  {title} {\bibinfo {title} {Layer-resolved quantum
  transport in twisted bilayer graphene: Counterflow and machine learning
  predictions},\ }\href {https://doi.org/10.1103/d98y-sv8j} {\bibfield
  {journal} {\bibinfo  {journal} {Physical Review B}\ }\textbf {\bibinfo
  {volume} {111}},\ \bibinfo {pages} {235425} (\bibinfo {year}
  {2025})}\BibitemShut {NoStop}%
\bibitem [{\citenamefont {Inui}\ \emph {et~al.}(2020)\citenamefont {Inui},
  \citenamefont {Aoki}, \citenamefont {Nishiue}, \citenamefont {Shiota},
  \citenamefont {Kousaka}, \citenamefont {Shishido}, \citenamefont {Hirobe},
  \citenamefont {Suda}, \citenamefont {Ohe}, \citenamefont {Kishine},
  \citenamefont {Yamamoto},\ and\ \citenamefont {Togawa}}]{InuiCrNb3S6}%
  \BibitemOpen
  \bibfield  {author} {\bibinfo {author} {\bibfnamefont {A.}~\bibnamefont
  {Inui}}, \bibinfo {author} {\bibfnamefont {R.}~\bibnamefont {Aoki}}, \bibinfo
  {author} {\bibfnamefont {Y.}~\bibnamefont {Nishiue}}, \bibinfo {author}
  {\bibfnamefont {K.}~\bibnamefont {Shiota}}, \bibinfo {author} {\bibfnamefont
  {Y.}~\bibnamefont {Kousaka}}, \bibinfo {author} {\bibfnamefont
  {H.}~\bibnamefont {Shishido}}, \bibinfo {author} {\bibfnamefont
  {D.}~\bibnamefont {Hirobe}}, \bibinfo {author} {\bibfnamefont
  {M.}~\bibnamefont {Suda}}, \bibinfo {author} {\bibfnamefont {J.-i.}\
  \bibnamefont {Ohe}}, \bibinfo {author} {\bibfnamefont {J.-i.}\ \bibnamefont
  {Kishine}}, \bibinfo {author} {\bibfnamefont {H.~M.}\ \bibnamefont
  {Yamamoto}},\ and\ \bibinfo {author} {\bibfnamefont {Y.}~\bibnamefont
  {Togawa}},\ }\bibfield  {title} {\bibinfo {title} {Chirality-induced
  spin-polarized state of a chiral crystal
  ${\mathrm{crnb}}_{3}{\mathrm{s}}_{6}$},\ }\href
  {https://doi.org/10.1103/PhysRevLett.124.166602} {\bibfield  {journal}
  {\bibinfo  {journal} {Physical Review Letters}\ }\textbf {\bibinfo {volume}
  {124}},\ \bibinfo {pages} {166602} (\bibinfo {year} {2020})}\BibitemShut
  {NoStop}%
\bibitem [{\citenamefont {Jo}\ \emph {et~al.}(2024)\citenamefont {Jo},
  \citenamefont {Go}, \citenamefont {Choi},\ and\ \citenamefont
  {Lee}}]{Rev_orbitronics}%
  \BibitemOpen
  \bibfield  {author} {\bibinfo {author} {\bibfnamefont {D.}~\bibnamefont
  {Jo}}, \bibinfo {author} {\bibfnamefont {D.}~\bibnamefont {Go}}, \bibinfo
  {author} {\bibfnamefont {G.-M.}\ \bibnamefont {Choi}},\ and\ \bibinfo
  {author} {\bibfnamefont {H.-W.}\ \bibnamefont {Lee}},\ }\bibfield  {title}
  {\bibinfo {title} {Spintronics meets orbitronics: Emergence of orbital
  angular momentum in solids},\ }\href
  {https://doi.org/10.1038/s44306-024-00023-6} {\bibfield  {journal} {\bibinfo
  {journal} {npj Spintronics}\ }\textbf {\bibinfo {volume} {2}},\ \bibinfo
  {pages} {19} (\bibinfo {year} {2024})}\BibitemShut {NoStop}%
\bibitem [{\citenamefont {Go}\ \emph {et~al.}(2021)\citenamefont {Go},
  \citenamefont {Jo}, \citenamefont {Lee}, \citenamefont {Kl{\"a}ui},\ and\
  \citenamefont {Mokrousov}}]{Go2021orbitaltorque}%
  \BibitemOpen
  \bibfield  {author} {\bibinfo {author} {\bibfnamefont {D.}~\bibnamefont
  {Go}}, \bibinfo {author} {\bibfnamefont {D.}~\bibnamefont {Jo}}, \bibinfo
  {author} {\bibfnamefont {H.-W.}\ \bibnamefont {Lee}}, \bibinfo {author}
  {\bibfnamefont {M.}~\bibnamefont {Kl{\"a}ui}},\ and\ \bibinfo {author}
  {\bibfnamefont {Y.}~\bibnamefont {Mokrousov}},\ }\bibfield  {title} {\bibinfo
  {title} {Orbitronics: Orbital currents in solids},\ }\href
  {https://doi.org/10.1209/0295-5075/ac2653} {\bibfield  {journal} {\bibinfo
  {journal} {Europhysics Letters}\ }\textbf {\bibinfo {volume} {135}},\
  \bibinfo {pages} {37001} (\bibinfo {year} {2021})}\BibitemShut {NoStop}%
\bibitem [{\citenamefont {Yoda}\ \emph {et~al.}(2018)\citenamefont {Yoda},
  \citenamefont {Yokoyama},\ and\ \citenamefont
  {Murakami}}]{doi:10.1021/acs.nanolett.7b04300}%
  \BibitemOpen
  \bibfield  {author} {\bibinfo {author} {\bibfnamefont {T.}~\bibnamefont
  {Yoda}}, \bibinfo {author} {\bibfnamefont {T.}~\bibnamefont {Yokoyama}},\
  and\ \bibinfo {author} {\bibfnamefont {S.}~\bibnamefont {Murakami}},\
  }\bibfield  {title} {\bibinfo {title} {Orbital edelstein effect as a
  condensed-matter analog of solenoids},\ }\href
  {https://doi.org/10.1021/acs.nanolett.7b04300} {\bibfield  {journal}
  {\bibinfo  {journal} {Nano Letters}\ }\textbf {\bibinfo {volume} {18}},\
  \bibinfo {pages} {916} (\bibinfo {year} {2018})}\BibitemShut {NoStop}%
\bibitem [{\citenamefont {Liu}\ \emph {et~al.}(2021{\natexlab{b}})\citenamefont
  {Liu}, \citenamefont {Xiao}, \citenamefont {Koo},\ and\ \citenamefont
  {Yan}}]{Binghai_DNA}%
  \BibitemOpen
  \bibfield  {author} {\bibinfo {author} {\bibfnamefont {Y.}~\bibnamefont
  {Liu}}, \bibinfo {author} {\bibfnamefont {J.}~\bibnamefont {Xiao}}, \bibinfo
  {author} {\bibfnamefont {J.}~\bibnamefont {Koo}},\ and\ \bibinfo {author}
  {\bibfnamefont {B.}~\bibnamefont {Yan}},\ }\bibfield  {title} {\bibinfo
  {title} {Chirality-driven topological electronic structure of dna-like
  materials},\ }\href {https://doi.org/10.1038/s41563-021-00924-5} {\bibfield
  {journal} {\bibinfo  {journal} {Nature Materials}\ }\textbf {\bibinfo
  {volume} {20}},\ \bibinfo {pages} {638} (\bibinfo {year}
  {2021}{\natexlab{b}})}\BibitemShut {NoStop}%
\bibitem [{\citenamefont {Kim}\ \emph {et~al.}(2023)\citenamefont {Kim},
  \citenamefont {Shin}, \citenamefont {Namgung}, \citenamefont {Park},
  \citenamefont {Kim},\ and\ \citenamefont
  {Kim}}]{doi:10.1021/acsnano.3c03893}%
  \BibitemOpen
  \bibfield  {author} {\bibinfo {author} {\bibfnamefont {B.}~\bibnamefont
  {Kim}}, \bibinfo {author} {\bibfnamefont {D.}~\bibnamefont {Shin}}, \bibinfo
  {author} {\bibfnamefont {S.}~\bibnamefont {Namgung}}, \bibinfo {author}
  {\bibfnamefont {N.}~\bibnamefont {Park}}, \bibinfo {author} {\bibfnamefont
  {K.-W.}\ \bibnamefont {Kim}},\ and\ \bibinfo {author} {\bibfnamefont
  {J.}~\bibnamefont {Kim}},\ }\bibfield  {title} {\bibinfo {title}
  {Optoelectronic manifestation of orbital angular momentum driven by chiral
  hopping in helical se chains},\ }\href
  {https://doi.org/10.1021/acsnano.3c03893} {\bibfield  {journal} {\bibinfo
  {journal} {ACS Nano}\ }\textbf {\bibinfo {volume} {17}},\ \bibinfo {pages}
  {18873} (\bibinfo {year} {2023})}\BibitemShut {NoStop}%
\bibitem [{\citenamefont {Hagiwara}\ \emph {et~al.}(2025)\citenamefont
  {Hagiwara}, \citenamefont {Chen}, \citenamefont {Go}, \citenamefont {Tan},
  \citenamefont {Grytsiuk}, \citenamefont {Yang}, \citenamefont {Shu},
  \citenamefont {Chien}, \citenamefont {Shen}, \citenamefont {Huang},
  \citenamefont {Cojocariu}, \citenamefont {Feyer}, \citenamefont {Lin},
  \citenamefont {Blügel}, \citenamefont {Schneider}, \citenamefont
  {Mokrousov},\ and\ \citenamefont
  {Tusche}}]{https://doi.org/10.1002/adma.202418040}%
  \BibitemOpen
  \bibfield  {author} {\bibinfo {author} {\bibfnamefont {K.}~\bibnamefont
  {Hagiwara}}, \bibinfo {author} {\bibfnamefont {Y.-J.}\ \bibnamefont {Chen}},
  \bibinfo {author} {\bibfnamefont {D.}~\bibnamefont {Go}}, \bibinfo {author}
  {\bibfnamefont {X.~L.}\ \bibnamefont {Tan}}, \bibinfo {author} {\bibfnamefont
  {S.}~\bibnamefont {Grytsiuk}}, \bibinfo {author} {\bibfnamefont {K.-H.~O.}\
  \bibnamefont {Yang}}, \bibinfo {author} {\bibfnamefont {G.-J.}\ \bibnamefont
  {Shu}}, \bibinfo {author} {\bibfnamefont {J.}~\bibnamefont {Chien}}, \bibinfo
  {author} {\bibfnamefont {Y.-H.}\ \bibnamefont {Shen}}, \bibinfo {author}
  {\bibfnamefont {X.-L.}\ \bibnamefont {Huang}}, \bibinfo {author}
  {\bibfnamefont {I.}~\bibnamefont {Cojocariu}}, \bibinfo {author}
  {\bibfnamefont {V.}~\bibnamefont {Feyer}}, \bibinfo {author} {\bibfnamefont
  {M.-T.}\ \bibnamefont {Lin}}, \bibinfo {author} {\bibfnamefont
  {S.}~\bibnamefont {Blügel}}, \bibinfo {author} {\bibfnamefont {C.~M.}\
  \bibnamefont {Schneider}}, \bibinfo {author} {\bibfnamefont {Y.}~\bibnamefont
  {Mokrousov}},\ and\ \bibinfo {author} {\bibfnamefont {C.}~\bibnamefont
  {Tusche}},\ }\bibfield  {title} {\bibinfo {title} {Orbital topology of chiral
  crystals for orbitronics},\ }\href
  {https://doi.org/https://doi.org/10.1002/adma.202418040} {\bibfield
  {journal} {\bibinfo  {journal} {Advanced Materials}\ }\textbf {\bibinfo
  {volume} {37}},\ \bibinfo {pages} {2418040} (\bibinfo {year}
  {2025})}\BibitemShut {NoStop}%
\bibitem [{\citenamefont {Bhowal}\ and\ \citenamefont
  {Vignale}(2021)}]{Bhowal2022}%
  \BibitemOpen
  \bibfield  {author} {\bibinfo {author} {\bibfnamefont {S.}~\bibnamefont
  {Bhowal}}\ and\ \bibinfo {author} {\bibfnamefont {G.}~\bibnamefont
  {Vignale}},\ }\bibfield  {title} {\bibinfo {title} {Orbital hall effect as an
  alternative to valley hall effect in gapped graphene},\ }\href
  {https://doi.org/10.1103/PhysRevB.103.195309} {\bibfield  {journal} {\bibinfo
   {journal} {Physical Review B}\ }\textbf {\bibinfo {volume} {103}},\ \bibinfo
  {pages} {195309} (\bibinfo {year} {2021})}\BibitemShut {NoStop}%
\bibitem [{\citenamefont {G\"obel}\ \emph {et~al.}(2025)\citenamefont
  {G\"obel}, \citenamefont {Schimpf},\ and\ \citenamefont
  {Mertig}}]{Gobel_CIOS}%
  \BibitemOpen
  \bibfield  {author} {\bibinfo {author} {\bibfnamefont {B.}~\bibnamefont
  {G\"obel}}, \bibinfo {author} {\bibfnamefont {L.}~\bibnamefont {Schimpf}},\
  and\ \bibinfo {author} {\bibfnamefont {I.}~\bibnamefont {Mertig}},\
  }\bibfield  {title} {\bibinfo {title} {Chirality-induced orbital edelstein
  effect in an analytically solvable model},\ }\href
  {https://doi.org/10.1103/vpjm-ntbh} {\bibfield  {journal} {\bibinfo
  {journal} {Phys. Rev. Res.}\ }\textbf {\bibinfo {volume} {7}},\ \bibinfo
  {pages} {033180} (\bibinfo {year} {2025})}\BibitemShut {NoStop}%
\bibitem [{\citenamefont {Salemi}\ \emph {et~al.}(2021)\citenamefont {Salemi},
  \citenamefont {Berritta},\ and\ \citenamefont {Oppeneer}}]{Salemi2022}%
  \BibitemOpen
  \bibfield  {author} {\bibinfo {author} {\bibfnamefont {L.}~\bibnamefont
  {Salemi}}, \bibinfo {author} {\bibfnamefont {M.}~\bibnamefont {Berritta}},\
  and\ \bibinfo {author} {\bibfnamefont {P.~M.}\ \bibnamefont {Oppeneer}},\
  }\bibfield  {title} {\bibinfo {title} {Quantitative comparison of
  electrically induced spin and orbital polarizations in heavy-metal/$3d$-metal
  bilayers},\ }\href {https://doi.org/10.1103/PhysRevMaterials.5.074407}
  {\bibfield  {journal} {\bibinfo  {journal} {Physical Review Materials}\
  }\textbf {\bibinfo {volume} {5}},\ \bibinfo {pages} {074407} (\bibinfo {year}
  {2021})}\BibitemShut {NoStop}%
\bibitem [{\citenamefont {G{\"o}bel}\ \emph {et~al.}(2025)\citenamefont
  {G{\"o}bel}, \citenamefont {Mertig},\ and\ \citenamefont
  {Lounis}}]{Gobel_CNT}%
  \BibitemOpen
  \bibfield  {author} {\bibinfo {author} {\bibfnamefont {B.}~\bibnamefont
  {G{\"o}bel}}, \bibinfo {author} {\bibfnamefont {I.}~\bibnamefont {Mertig}},\
  and\ \bibinfo {author} {\bibfnamefont {S.}~\bibnamefont {Lounis}},\
  }\bibfield  {title} {\bibinfo {title} {Chirality-induced selectivity of
  angular momentum by orbital edelstein effect in carbon nanotubes},\ }\href
  {https://doi.org/10.1038/s42005-025-02331-7} {\bibfield  {journal} {\bibinfo
  {journal} {Communications Physics}\ }\textbf {\bibinfo {volume} {8}},\
  \bibinfo {pages} {395} (\bibinfo {year} {2025})}\BibitemShut {NoStop}%
\bibitem [{\citenamefont {Charlier}\ \emph {et~al.}(2007)\citenamefont
  {Charlier}, \citenamefont {Blase},\ and\ \citenamefont
  {Roche}}]{RevModPhys.79.677}%
  \BibitemOpen
  \bibfield  {author} {\bibinfo {author} {\bibfnamefont {J.-C.}\ \bibnamefont
  {Charlier}}, \bibinfo {author} {\bibfnamefont {X.}~\bibnamefont {Blase}},\
  and\ \bibinfo {author} {\bibfnamefont {S.}~\bibnamefont {Roche}},\ }\bibfield
   {title} {\bibinfo {title} {Electronic and transport properties of
  nanotubes},\ }\href {https://doi.org/10.1103/RevModPhys.79.677} {\bibfield
  {journal} {\bibinfo  {journal} {Reviews of Modern Physics}\ }\textbf
  {\bibinfo {volume} {79}},\ \bibinfo {pages} {677} (\bibinfo {year}
  {2007})}\BibitemShut {NoStop}%
\bibitem [{\citenamefont {Laird}\ \emph {et~al.}(2015)\citenamefont {Laird},
  \citenamefont {Kuemmeth}, \citenamefont {Steele}, \citenamefont
  {Grove-Rasmussen}, \citenamefont {Nyg\aa{}rd}, \citenamefont {Flensberg},\
  and\ \citenamefont {Kouwenhoven}}]{RevModPhys.87.703}%
  \BibitemOpen
  \bibfield  {author} {\bibinfo {author} {\bibfnamefont {E.~A.}\ \bibnamefont
  {Laird}}, \bibinfo {author} {\bibfnamefont {F.}~\bibnamefont {Kuemmeth}},
  \bibinfo {author} {\bibfnamefont {G.~A.}\ \bibnamefont {Steele}}, \bibinfo
  {author} {\bibfnamefont {K.}~\bibnamefont {Grove-Rasmussen}}, \bibinfo
  {author} {\bibfnamefont {J.}~\bibnamefont {Nyg\aa{}rd}}, \bibinfo {author}
  {\bibfnamefont {K.}~\bibnamefont {Flensberg}},\ and\ \bibinfo {author}
  {\bibfnamefont {L.~P.}\ \bibnamefont {Kouwenhoven}},\ }\bibfield  {title}
  {\bibinfo {title} {Quantum transport in carbon nanotubes},\ }\href
  {https://doi.org/10.1103/RevModPhys.87.703} {\bibfield  {journal} {\bibinfo
  {journal} {Reviews of Modern Physics}\ }\textbf {\bibinfo {volume} {87}},\
  \bibinfo {pages} {703} (\bibinfo {year} {2015})}\BibitemShut {NoStop}%
\bibitem [{\citenamefont {Minot}\ \emph
  {et~al.}(2004{\natexlab{a}})\citenamefont {Minot}, \citenamefont {Yaish},
  \citenamefont {Sazonova},\ and\ \citenamefont {McEuen}}]{Minot2004}%
  \BibitemOpen
  \bibfield  {author} {\bibinfo {author} {\bibfnamefont {E.~D.}\ \bibnamefont
  {Minot}}, \bibinfo {author} {\bibfnamefont {Y.}~\bibnamefont {Yaish}},
  \bibinfo {author} {\bibfnamefont {V.}~\bibnamefont {Sazonova}},\ and\
  \bibinfo {author} {\bibfnamefont {P.~L.}\ \bibnamefont {McEuen}},\ }\bibfield
   {title} {\bibinfo {title} {Determination of electron orbital magnetic
  moments in carbon nanotubes},\ }\href {https://doi.org/10.1038/nature02425}
  {\bibfield  {journal} {\bibinfo  {journal} {Nature}\ }\textbf {\bibinfo
  {volume} {428}},\ \bibinfo {pages} {536} (\bibinfo {year}
  {2004}{\natexlab{a}})}\BibitemShut {NoStop}%
\bibitem [{\citenamefont {Jarillo-Herrero}\ \emph {et~al.}(2005)\citenamefont
  {Jarillo-Herrero}, \citenamefont {Kong}, \citenamefont {van~der Zant},
  \citenamefont {Dekker}, \citenamefont {Kouwenhoven},\ and\ \citenamefont
  {De~Franceschi}}]{Jarillo2005}%
  \BibitemOpen
  \bibfield  {author} {\bibinfo {author} {\bibfnamefont {P.}~\bibnamefont
  {Jarillo-Herrero}}, \bibinfo {author} {\bibfnamefont {J.}~\bibnamefont
  {Kong}}, \bibinfo {author} {\bibfnamefont {H.~S.~J.}\ \bibnamefont {van~der
  Zant}}, \bibinfo {author} {\bibfnamefont {C.}~\bibnamefont {Dekker}},
  \bibinfo {author} {\bibfnamefont {L.~P.}\ \bibnamefont {Kouwenhoven}},\ and\
  \bibinfo {author} {\bibfnamefont {S.}~\bibnamefont {De~Franceschi}},\
  }\bibfield  {title} {\bibinfo {title} {Orbital kondo effect in carbon
  nanotubes},\ }\href {https://doi.org/10.1038/nature03422} {\bibfield
  {journal} {\bibinfo  {journal} {Nature}\ }\textbf {\bibinfo {volume} {434}},\
  \bibinfo {pages} {484} (\bibinfo {year} {2005})}\BibitemShut {NoStop}%
\bibitem [{\citenamefont {Miyamoto}\ \emph {et~al.}(1996)\citenamefont
  {Miyamoto}, \citenamefont {Louie},\ and\ \citenamefont
  {Cohen}}]{PhysRevLett.76.2121}%
  \BibitemOpen
  \bibfield  {author} {\bibinfo {author} {\bibfnamefont {Y.}~\bibnamefont
  {Miyamoto}}, \bibinfo {author} {\bibfnamefont {S.~G.}\ \bibnamefont
  {Louie}},\ and\ \bibinfo {author} {\bibfnamefont {M.~L.}\ \bibnamefont
  {Cohen}},\ }\bibfield  {title} {\bibinfo {title} {Chiral conductivities of
  nanotubes},\ }\href {https://doi.org/10.1103/PhysRevLett.76.2121} {\bibfield
  {journal} {\bibinfo  {journal} {Physical Review Letters}\ }\textbf {\bibinfo
  {volume} {76}},\ \bibinfo {pages} {2121} (\bibinfo {year}
  {1996})}\BibitemShut {NoStop}%
\bibitem [{\citenamefont {Miyamoto}\ \emph {et~al.}(1999)\citenamefont
  {Miyamoto}, \citenamefont {Rubio}, \citenamefont {Louie},\ and\ \citenamefont
  {Cohen}}]{PhysRevB.60.13885}%
  \BibitemOpen
  \bibfield  {author} {\bibinfo {author} {\bibfnamefont {Y.}~\bibnamefont
  {Miyamoto}}, \bibinfo {author} {\bibfnamefont {A.}~\bibnamefont {Rubio}},
  \bibinfo {author} {\bibfnamefont {S.~G.}\ \bibnamefont {Louie}},\ and\
  \bibinfo {author} {\bibfnamefont {M.~L.}\ \bibnamefont {Cohen}},\ }\bibfield
  {title} {\bibinfo {title} {Self-inductance of chiral conducting nanotubes},\
  }\href {https://doi.org/10.1103/PhysRevB.60.13885} {\bibfield  {journal}
  {\bibinfo  {journal} {Physical Review B}\ }\textbf {\bibinfo {volume} {60}},\
  \bibinfo {pages} {13885} (\bibinfo {year} {1999})}\BibitemShut {NoStop}%
\bibitem [{\citenamefont {Lambert}\ \emph {et~al.}(2008)\citenamefont
  {Lambert}, \citenamefont {Bailey},\ and\ \citenamefont
  {Cserti}}]{PhysRevB.78.233405}%
  \BibitemOpen
  \bibfield  {author} {\bibinfo {author} {\bibfnamefont {C.~J.}\ \bibnamefont
  {Lambert}}, \bibinfo {author} {\bibfnamefont {S.~W.~D.}\ \bibnamefont
  {Bailey}},\ and\ \bibinfo {author} {\bibfnamefont {J.}~\bibnamefont
  {Cserti}},\ }\bibfield  {title} {\bibinfo {title} {Oscillating chiral
  currents in nanotubes: A route to nanoscale magnetic test tubes},\ }\href
  {https://doi.org/10.1103/PhysRevB.78.233405} {\bibfield  {journal} {\bibinfo
  {journal} {Physical Review B}\ }\textbf {\bibinfo {volume} {78}},\ \bibinfo
  {pages} {233405} (\bibinfo {year} {2008})}\BibitemShut {NoStop}%
\bibitem [{\citenamefont {Marganska}\ \emph {et~al.}(2015)\citenamefont
  {Marganska}, \citenamefont {Chudzinski},\ and\ \citenamefont
  {Grifoni}}]{PhysRevB.92.075433}%
  \BibitemOpen
  \bibfield  {author} {\bibinfo {author} {\bibfnamefont {M.}~\bibnamefont
  {Marganska}}, \bibinfo {author} {\bibfnamefont {P.}~\bibnamefont
  {Chudzinski}},\ and\ \bibinfo {author} {\bibfnamefont {M.}~\bibnamefont
  {Grifoni}},\ }\bibfield  {title} {\bibinfo {title} {The two classes of
  low-energy spectra in finite carbon nanotubes},\ }\href
  {https://doi.org/10.1103/PhysRevB.92.075433} {\bibfield  {journal} {\bibinfo
  {journal} {Physical Review B}\ }\textbf {\bibinfo {volume} {92}},\ \bibinfo
  {pages} {075433} (\bibinfo {year} {2015})}\BibitemShut {NoStop}%
\bibitem [{\citenamefont {Izumida}\ \emph {et~al.}(2015)\citenamefont
  {Izumida}, \citenamefont {Okuyama},\ and\ \citenamefont
  {Saito}}]{PhysRevB.91.235442}%
  \BibitemOpen
  \bibfield  {author} {\bibinfo {author} {\bibfnamefont {W.}~\bibnamefont
  {Izumida}}, \bibinfo {author} {\bibfnamefont {R.}~\bibnamefont {Okuyama}},\
  and\ \bibinfo {author} {\bibfnamefont {R.}~\bibnamefont {Saito}},\ }\bibfield
   {title} {\bibinfo {title} {Valley coupling in finite-length metallic
  single-wall carbon nanotubes},\ }\href
  {https://doi.org/10.1103/PhysRevB.91.235442} {\bibfield  {journal} {\bibinfo
  {journal} {Physical Review B}\ }\textbf {\bibinfo {volume} {91}},\ \bibinfo
  {pages} {235442} (\bibinfo {year} {2015})}\BibitemShut {NoStop}%
\bibitem [{\citenamefont {Izumida}\ \emph {et~al.}(2016)\citenamefont
  {Izumida}, \citenamefont {Okuyama}, \citenamefont {Yamakage},\ and\
  \citenamefont {Saito}}]{PhysRevB.93.195442}%
  \BibitemOpen
  \bibfield  {author} {\bibinfo {author} {\bibfnamefont {W.}~\bibnamefont
  {Izumida}}, \bibinfo {author} {\bibfnamefont {R.}~\bibnamefont {Okuyama}},
  \bibinfo {author} {\bibfnamefont {A.}~\bibnamefont {Yamakage}},\ and\
  \bibinfo {author} {\bibfnamefont {R.}~\bibnamefont {Saito}},\ }\bibfield
  {title} {\bibinfo {title} {Angular momentum and topology in semiconducting
  single-wall carbon nanotubes},\ }\href
  {https://doi.org/10.1103/PhysRevB.93.195442} {\bibfield  {journal} {\bibinfo
  {journal} {Physical Review B}\ }\textbf {\bibinfo {volume} {93}},\ \bibinfo
  {pages} {195442} (\bibinfo {year} {2016})}\BibitemShut {NoStop}%
\bibitem [{\citenamefont {Marga\ifmmode~\acute{n}\else \'{n}\fi{}ska}\ \emph
  {et~al.}(2019)\citenamefont {Marga\ifmmode~\acute{n}\else \'{n}\fi{}ska},
  \citenamefont {Schmid}, \citenamefont {Dirnaichner}, \citenamefont {Stiller},
  \citenamefont {Strunk}, \citenamefont {Grifoni},\ and\ \citenamefont
  {H\"uttel}}]{Marganska2019}%
  \BibitemOpen
  \bibfield  {author} {\bibinfo {author} {\bibfnamefont {M.}~\bibnamefont
  {Marga\ifmmode~\acute{n}\else \'{n}\fi{}ska}}, \bibinfo {author}
  {\bibfnamefont {D.~R.}\ \bibnamefont {Schmid}}, \bibinfo {author}
  {\bibfnamefont {A.}~\bibnamefont {Dirnaichner}}, \bibinfo {author}
  {\bibfnamefont {P.~L.}\ \bibnamefont {Stiller}}, \bibinfo {author}
  {\bibfnamefont {C.}~\bibnamefont {Strunk}}, \bibinfo {author} {\bibfnamefont
  {M.}~\bibnamefont {Grifoni}},\ and\ \bibinfo {author} {\bibfnamefont {A.~K.}\
  \bibnamefont {H\"uttel}},\ }\bibfield  {title} {\bibinfo {title} {Shaping
  electron wave functions in a carbon nanotube with a parallel magnetic
  field},\ }\href {https://doi.org/10.1103/PhysRevLett.122.086802} {\bibfield
  {journal} {\bibinfo  {journal} {Physical Review Letters}\ }\textbf {\bibinfo
  {volume} {122}},\ \bibinfo {pages} {086802} (\bibinfo {year}
  {2019})}\BibitemShut {NoStop}%
\bibitem [{\citenamefont {Frey}\ and\ \citenamefont
  {Doren}(2011)}]{tubegen2011}%
  \BibitemOpen
  \bibfield  {author} {\bibinfo {author} {\bibfnamefont {J.~T.}\ \bibnamefont
  {Frey}}\ and\ \bibinfo {author} {\bibfnamefont {D.~J.}\ \bibnamefont
  {Doren}},\ }\href@noop {} {\bibinfo {title} {Tubegen 3.4}},\ \bibinfo
  {howpublished} {\url{http://turin.nss.udel.edu/research/tubegenonline.html}}
  (\bibinfo {year} {2011}),\ \bibinfo {note} {web interface, University of
  Delaware, Newark, DE}\BibitemShut {NoStop}%
\bibitem [{\citenamefont {Saito}\ \emph {et~al.}(1998)\citenamefont {Saito},
  \citenamefont {Dresselhaus},\ and\ \citenamefont {Dresselhaus}}]{bookSaito}%
  \BibitemOpen
  \bibfield  {author} {\bibinfo {author} {\bibfnamefont {R.}~\bibnamefont
  {Saito}}, \bibinfo {author} {\bibfnamefont {G.}~\bibnamefont {Dresselhaus}},\
  and\ \bibinfo {author} {\bibfnamefont {M.~S.}\ \bibnamefont {Dresselhaus}},\
  }\href@noop {} {\emph {\bibinfo {title} {Physical Properties of Carbon
  Nanotubes}}}\ (\bibinfo  {publisher} {Imperial College Press},\ \bibinfo
  {address} {London},\ \bibinfo {year} {1998})\BibitemShut {NoStop}%
\bibitem [{\citenamefont {Dresselhaus}\ \emph {et~al.}(2008)\citenamefont
  {Dresselhaus}, \citenamefont {Dresselhaus}, \citenamefont {Saito},\ and\
  \citenamefont {Jorio}}]{DresselhausBook}%
  \BibitemOpen
  \bibfield  {author} {\bibinfo {author} {\bibfnamefont {M.~S.}\ \bibnamefont
  {Dresselhaus}}, \bibinfo {author} {\bibfnamefont {G.}~\bibnamefont
  {Dresselhaus}}, \bibinfo {author} {\bibfnamefont {R.}~\bibnamefont {Saito}},\
  and\ \bibinfo {author} {\bibfnamefont {A.}~\bibnamefont {Jorio}},\ }\href
  {https://doi.org/10.1007/978-3-540-32899-5} {\emph {\bibinfo {title} {Group
  Theory: Application to the Physics of Condensed Matter}}}\ (\bibinfo
  {publisher} {Springer-Verlag},\ \bibinfo {address} {Berlin Heidelberg},\
  \bibinfo {year} {2008})\BibitemShut {NoStop}%
\bibitem [{\citenamefont {White}\ and\ \citenamefont
  {Mintmire}(2005)}]{WhiteMintmire1998}%
  \BibitemOpen
  \bibfield  {author} {\bibinfo {author} {\bibfnamefont {C.~T.}\ \bibnamefont
  {White}}\ and\ \bibinfo {author} {\bibfnamefont {J.~W.}\ \bibnamefont
  {Mintmire}},\ }\bibfield  {title} {\bibinfo {title} {Fundamental properties
  of single-wall carbon nanotubes},\ }\href {https://doi.org/10.1021/jp047416+}
  {\bibfield  {journal} {\bibinfo  {journal} {The Journal of Physical Chemistry
  B}\ }\textbf {\bibinfo {volume} {109}},\ \bibinfo {pages} {52} (\bibinfo
  {year} {2005})}\BibitemShut {NoStop}%
\bibitem [{\citenamefont {Reich}\ \emph
  {et~al.}(2004{\natexlab{a}})\citenamefont {Reich}, \citenamefont {Thomsen},\
  and\ \citenamefont {Maultzsch}}]{Reich2004book}%
  \BibitemOpen
  \bibfield  {author} {\bibinfo {author} {\bibfnamefont {S.}~\bibnamefont
  {Reich}}, \bibinfo {author} {\bibfnamefont {C.}~\bibnamefont {Thomsen}},\
  and\ \bibinfo {author} {\bibfnamefont {J.}~\bibnamefont {Maultzsch}},\ }\href
  {https://doi.org/10.1002/9783527618040} {\emph {\bibinfo {title} {Carbon
  Nanotubes: Basic Concepts and Physical Properties}}}\ (\bibinfo  {publisher}
  {Wiley-VCH Verlag GmbH \& Co. KGaA},\ \bibinfo {address} {Weinheim,
  Germany},\ \bibinfo {year} {2004})\BibitemShut {NoStop}%
\bibitem [{\citenamefont {Datta}(1995)}]{Datta1995}%
  \BibitemOpen
  \bibfield  {author} {\bibinfo {author} {\bibfnamefont {S.}~\bibnamefont
  {Datta}},\ }\href {https://doi.org/10.1017/CBO9780511805776} {\emph {\bibinfo
  {title} {Electronic Transport in Mesoscopic Systems}}}\ (\bibinfo
  {publisher} {Cambridge University Press},\ \bibinfo {address} {Cambridge},\
  \bibinfo {year} {1995})\BibitemShut {NoStop}%
\bibitem [{\citenamefont {Bahamon}\ \emph {et~al.}(2020)\citenamefont
  {Bahamon}, \citenamefont {Gómez-Santos},\ and\ \citenamefont
  {Stauber}}]{bahamon2020emergent}%
  \BibitemOpen
  \bibfield  {author} {\bibinfo {author} {\bibfnamefont {D.~A.}\ \bibnamefont
  {Bahamon}}, \bibinfo {author} {\bibfnamefont {G.}~\bibnamefont
  {Gómez-Santos}},\ and\ \bibinfo {author} {\bibfnamefont {T.}~\bibnamefont
  {Stauber}},\ }\bibfield  {title} {\bibinfo {title} {Emergent magnetic texture
  in driven twisted bilayer graphene},\ }\href
  {https://doi.org/10.1039/D0NR02786C} {\bibfield  {journal} {\bibinfo
  {journal} {Nanoscale}\ }\textbf {\bibinfo {volume} {12}},\ \bibinfo {pages}
  {15383} (\bibinfo {year} {2020})}\BibitemShut {NoStop}%
\bibitem [{\citenamefont {Haug}\ and\ \citenamefont
  {Jauho}(2008)}]{HaugJauho2008}%
  \BibitemOpen
  \bibfield  {author} {\bibinfo {author} {\bibfnamefont {H.}~\bibnamefont
  {Haug}}\ and\ \bibinfo {author} {\bibfnamefont {A.-P.}\ \bibnamefont
  {Jauho}},\ }\href {https://doi.org/10.1007/978-3-540-73564-9} {\emph
  {\bibinfo {title} {Quantum Kinetics in Transport and Optics of
  Semiconductors}}},\ \bibinfo {edition} {2nd}\ ed.\ (\bibinfo  {publisher}
  {Springer-Verlag},\ \bibinfo {address} {Berlin Heidelberg},\ \bibinfo {year}
  {2008})\BibitemShut {NoStop}%
\bibitem [{\citenamefont {Peierls}(1933)}]{Peierls1933}%
  \BibitemOpen
  \bibfield  {author} {\bibinfo {author} {\bibfnamefont {R.}~\bibnamefont
  {Peierls}},\ }\bibfield  {title} {\bibinfo {title} {{Zur Theorie des
  Diamagnetismus von Leitungselektronen}},\ }\href
  {https://doi.org/10.1007/BF01342591} {\bibfield  {journal} {\bibinfo
  {journal} {Zeitschrift f{\"u}r Physik}\ }\textbf {\bibinfo {volume} {80}},\
  \bibinfo {pages} {763} (\bibinfo {year} {1933})}\BibitemShut {NoStop}%
\bibitem [{\citenamefont {Bahamon}\ \emph {et~al.}(2013)\citenamefont
  {Bahamon}, \citenamefont {Castro~Neto},\ and\ \citenamefont
  {Pereira}}]{Bahamon}%
  \BibitemOpen
  \bibfield  {author} {\bibinfo {author} {\bibfnamefont {D.~A.}\ \bibnamefont
  {Bahamon}}, \bibinfo {author} {\bibfnamefont {A.~H.}\ \bibnamefont
  {Castro~Neto}},\ and\ \bibinfo {author} {\bibfnamefont {V.~M.}\ \bibnamefont
  {Pereira}},\ }\bibfield  {title} {\bibinfo {title} {Effective contact model
  for geometry-independent conductance calculations in graphene},\ }\href
  {https://doi.org/10.1103/PhysRevB.88.235433} {\bibfield  {journal} {\bibinfo
  {journal} {Physical Review B}\ }\textbf {\bibinfo {volume} {88}},\ \bibinfo
  {pages} {235433} (\bibinfo {year} {2013})}\BibitemShut {NoStop}%
\bibitem [{\citenamefont {Nemec}\ \emph {et~al.}(2006)\citenamefont {Nemec},
  \citenamefont {Tom\'anek},\ and\ \citenamefont {Cuniberti}}]{Nemec2006}%
  \BibitemOpen
  \bibfield  {author} {\bibinfo {author} {\bibfnamefont {N.}~\bibnamefont
  {Nemec}}, \bibinfo {author} {\bibfnamefont {D.}~\bibnamefont {Tom\'anek}},\
  and\ \bibinfo {author} {\bibfnamefont {G.}~\bibnamefont {Cuniberti}},\
  }\bibfield  {title} {\bibinfo {title} {Contact dependence of carrier
  injection in carbon nanotubes: An ab initio study},\ }\href
  {https://doi.org/10.1103/PhysRevLett.96.076802} {\bibfield  {journal}
  {\bibinfo  {journal} {Physical Review Letters}\ }\textbf {\bibinfo {volume}
  {96}},\ \bibinfo {pages} {076802} (\bibinfo {year} {2006})}\BibitemShut
  {NoStop}%
\bibitem [{\citenamefont {Tersoff}(1999)}]{Tersoff1999}%
  \BibitemOpen
  \bibfield  {author} {\bibinfo {author} {\bibfnamefont {J.}~\bibnamefont
  {Tersoff}},\ }\bibfield  {title} {\bibinfo {title} {Contact resistance of
  carbon nanotubes},\ }\href {https://doi.org/10.1063/1.123776} {\bibfield
  {journal} {\bibinfo  {journal} {Applied Physics Letters}\ }\textbf {\bibinfo
  {volume} {74}},\ \bibinfo {pages} {2122} (\bibinfo {year}
  {1999})}\BibitemShut {NoStop}%
\bibitem [{\citenamefont {Chen}\ \emph {et~al.}(2005)\citenamefont {Chen},
  \citenamefont {Appenzeller}, \citenamefont {Knoch}, \citenamefont {Lin},\
  and\ \citenamefont {Avouris}}]{Chen2005contact}%
  \BibitemOpen
  \bibfield  {author} {\bibinfo {author} {\bibfnamefont {Z.}~\bibnamefont
  {Chen}}, \bibinfo {author} {\bibfnamefont {J.}~\bibnamefont {Appenzeller}},
  \bibinfo {author} {\bibfnamefont {J.}~\bibnamefont {Knoch}}, \bibinfo
  {author} {\bibfnamefont {Y.-m.}\ \bibnamefont {Lin}},\ and\ \bibinfo {author}
  {\bibfnamefont {P.}~\bibnamefont {Avouris}},\ }\bibfield  {title} {\bibinfo
  {title} {The role of metal--nanotube contact in the performance of carbon
  nanotube field-effect transistors},\ }\href
  {https://doi.org/10.1021/nl0508624} {\bibfield  {journal} {\bibinfo
  {journal} {Nano Letters}\ }\textbf {\bibinfo {volume} {5}},\ \bibinfo {pages}
  {1497} (\bibinfo {year} {2005})}\BibitemShut {NoStop}%
\bibitem [{\citenamefont {Zhang}\ \emph {et~al.}(2017)\citenamefont {Zhang},
  \citenamefont {Kang}, \citenamefont {Wang}, \citenamefont {Tong},
  \citenamefont {Yang}, \citenamefont {Wang}, \citenamefont {Qi}, \citenamefont
  {Deng}, \citenamefont {Li}, \citenamefont {Bai}, \citenamefont {Ding},\ and\
  \citenamefont {Zhang}}]{Zhang2017CVD}%
  \BibitemOpen
  \bibfield  {author} {\bibinfo {author} {\bibfnamefont {S.}~\bibnamefont
  {Zhang}}, \bibinfo {author} {\bibfnamefont {L.}~\bibnamefont {Kang}},
  \bibinfo {author} {\bibfnamefont {X.}~\bibnamefont {Wang}}, \bibinfo {author}
  {\bibfnamefont {L.}~\bibnamefont {Tong}}, \bibinfo {author} {\bibfnamefont
  {L.}~\bibnamefont {Yang}}, \bibinfo {author} {\bibfnamefont {Z.}~\bibnamefont
  {Wang}}, \bibinfo {author} {\bibfnamefont {K.}~\bibnamefont {Qi}}, \bibinfo
  {author} {\bibfnamefont {S.}~\bibnamefont {Deng}}, \bibinfo {author}
  {\bibfnamefont {Q.}~\bibnamefont {Li}}, \bibinfo {author} {\bibfnamefont
  {X.}~\bibnamefont {Bai}}, \bibinfo {author} {\bibfnamefont {F.}~\bibnamefont
  {Ding}},\ and\ \bibinfo {author} {\bibfnamefont {J.}~\bibnamefont {Zhang}},\
  }\bibfield  {title} {\bibinfo {title} {Arrays of horizontal carbon nanotubes
  of controlled chirality grown using designed catalysts},\ }\href
  {https://doi.org/10.1038/nature21051} {\bibfield  {journal} {\bibinfo
  {journal} {Nature}\ }\textbf {\bibinfo {volume} {543}},\ \bibinfo {pages}
  {234} (\bibinfo {year} {2017})}\BibitemShut {NoStop}%
\bibitem [{\citenamefont {Blase}\ \emph {et~al.}(1994)\citenamefont {Blase},
  \citenamefont {Benedict}, \citenamefont {Shirley},\ and\ \citenamefont
  {Louie}}]{Blase1994}%
  \BibitemOpen
  \bibfield  {author} {\bibinfo {author} {\bibfnamefont {X.}~\bibnamefont
  {Blase}}, \bibinfo {author} {\bibfnamefont {L.~X.}\ \bibnamefont {Benedict}},
  \bibinfo {author} {\bibfnamefont {E.~L.}\ \bibnamefont {Shirley}},\ and\
  \bibinfo {author} {\bibfnamefont {S.~G.}\ \bibnamefont {Louie}},\ }\bibfield
  {title} {\bibinfo {title} {Hybridization effects and metallicity in small
  radius carbon nanotubes},\ }\href
  {https://doi.org/10.1103/PhysRevLett.72.1878} {\bibfield  {journal} {\bibinfo
   {journal} {Physical Review Letters}\ }\textbf {\bibinfo {volume} {72}},\
  \bibinfo {pages} {1878} (\bibinfo {year} {1994})}\BibitemShut {NoStop}%
\bibitem [{\citenamefont {Salemi}\ and\ \citenamefont
  {Oppeneer}(2022)}]{Salemi2022orbital}%
  \BibitemOpen
  \bibfield  {author} {\bibinfo {author} {\bibfnamefont {L.}~\bibnamefont
  {Salemi}}\ and\ \bibinfo {author} {\bibfnamefont {P.~M.}\ \bibnamefont
  {Oppeneer}},\ }\bibfield  {title} {\bibinfo {title} {First-principles theory
  of intrinsic spin and orbital hall and nernst effects in metallic monoatomic
  crystals},\ }\href {https://doi.org/10.1103/PhysRevMaterials.6.095001}
  {\bibfield  {journal} {\bibinfo  {journal} {Physical Review Materials}\
  }\textbf {\bibinfo {volume} {6}},\ \bibinfo {pages} {095001} (\bibinfo {year}
  {2022})}\BibitemShut {NoStop}%
\bibitem [{\citenamefont {Kataura}\ \emph {et~al.}(1999)\citenamefont
  {Kataura}, \citenamefont {Kumazawa}, \citenamefont {Maniwa}, \citenamefont
  {Umezu}, \citenamefont {Suzuki}, \citenamefont {Ohtsuka},\ and\ \citenamefont
  {Achiba}}]{KATAURA19992555}%
  \BibitemOpen
  \bibfield  {author} {\bibinfo {author} {\bibfnamefont {H.}~\bibnamefont
  {Kataura}}, \bibinfo {author} {\bibfnamefont {Y.}~\bibnamefont {Kumazawa}},
  \bibinfo {author} {\bibfnamefont {Y.}~\bibnamefont {Maniwa}}, \bibinfo
  {author} {\bibfnamefont {I.}~\bibnamefont {Umezu}}, \bibinfo {author}
  {\bibfnamefont {S.}~\bibnamefont {Suzuki}}, \bibinfo {author} {\bibfnamefont
  {Y.}~\bibnamefont {Ohtsuka}},\ and\ \bibinfo {author} {\bibfnamefont
  {Y.}~\bibnamefont {Achiba}},\ }\bibfield  {title} {\bibinfo {title} {Optical
  properties of single-wall carbon nanotubes},\ }\href
  {https://doi.org/https://doi.org/10.1016/S0379-6779(98)00278-1} {\bibfield
  {journal} {\bibinfo  {journal} {Synthetic Metals}\ }\textbf {\bibinfo
  {volume} {103}},\ \bibinfo {pages} {2555} (\bibinfo {year} {1999})},\
  \bibinfo {note} {international Conference on Science and Technology of
  Synthetic Metals}\BibitemShut {NoStop}%
\bibitem [{\citenamefont {Reich}\ \emph
  {et~al.}(2004{\natexlab{b}})\citenamefont {Reich}, \citenamefont {Thomsen},\
  and\ \citenamefont {Maultzsch}}]{Reich2004}%
  \BibitemOpen
  \bibfield  {author} {\bibinfo {author} {\bibfnamefont {S.}~\bibnamefont
  {Reich}}, \bibinfo {author} {\bibfnamefont {C.}~\bibnamefont {Thomsen}},\
  and\ \bibinfo {author} {\bibfnamefont {J.}~\bibnamefont {Maultzsch}},\
  }\href@noop {} {\emph {\bibinfo {title} {Carbon Nanotubes: Basic Concepts and
  Physical Properties}}}\ (\bibinfo  {publisher} {John Wiley \& Sons, Ltd},\
  \bibinfo {address} {Weinheim, Germany},\ \bibinfo {year} {2004})\
  Chap.~\bibinfo {chapter} {4}, pp.\ \bibinfo {pages} {67--83}\BibitemShut
  {NoStop}%
\bibitem [{\citenamefont {Minot}\ \emph
  {et~al.}(2004{\natexlab{b}})\citenamefont {Minot}, \citenamefont {Yaish},
  \citenamefont {Sazonova},\ and\ \citenamefont {McEuen}}]{McEuen_muexp}%
  \BibitemOpen
  \bibfield  {author} {\bibinfo {author} {\bibfnamefont {E.~D.}\ \bibnamefont
  {Minot}}, \bibinfo {author} {\bibfnamefont {Y.}~\bibnamefont {Yaish}},
  \bibinfo {author} {\bibfnamefont {V.}~\bibnamefont {Sazonova}},\ and\
  \bibinfo {author} {\bibfnamefont {P.~L.}\ \bibnamefont {McEuen}},\ }\bibfield
   {title} {\bibinfo {title} {Determination of electron orbital magnetic
  moments in carbon nanotubes},\ }\href {https://doi.org/10.1038/nature02425}
  {\bibfield  {journal} {\bibinfo  {journal} {Nature}\ }\textbf {\bibinfo
  {volume} {428}},\ \bibinfo {pages} {536} (\bibinfo {year}
  {2004}{\natexlab{b}})}\BibitemShut {NoStop}%
\bibitem [{\citenamefont {Ajiki}\ and\ \citenamefont {Ando}(1993)}]{Ajiki1993}%
  \BibitemOpen
  \bibfield  {author} {\bibinfo {author} {\bibfnamefont {H.}~\bibnamefont
  {Ajiki}}\ and\ \bibinfo {author} {\bibfnamefont {T.}~\bibnamefont {Ando}},\
  }\bibfield  {title} {\bibinfo {title} {Magnetic properties of carbon
  nanotubes},\ }\href {https://doi.org/10.1143/JPSJ.62.2470} {\bibfield
  {journal} {\bibinfo  {journal} {Journal of the Physical Society of Japan}\
  }\textbf {\bibinfo {volume} {62}},\ \bibinfo {pages} {2470} (\bibinfo {year}
  {1993})}\BibitemShut {NoStop}%
\bibitem [{\citenamefont {Bachilo}\ \emph {et~al.}(2002)\citenamefont
  {Bachilo}, \citenamefont {Strano}, \citenamefont {Kittrell}, \citenamefont
  {Hauge}, \citenamefont {Smalley},\ and\ \citenamefont
  {Weisman}}]{Bachilo2002}%
  \BibitemOpen
  \bibfield  {author} {\bibinfo {author} {\bibfnamefont {S.~M.}\ \bibnamefont
  {Bachilo}}, \bibinfo {author} {\bibfnamefont {M.~S.}\ \bibnamefont {Strano}},
  \bibinfo {author} {\bibfnamefont {C.}~\bibnamefont {Kittrell}}, \bibinfo
  {author} {\bibfnamefont {R.~H.}\ \bibnamefont {Hauge}}, \bibinfo {author}
  {\bibfnamefont {R.~E.}\ \bibnamefont {Smalley}},\ and\ \bibinfo {author}
  {\bibfnamefont {R.~B.}\ \bibnamefont {Weisman}},\ }\bibfield  {title}
  {\bibinfo {title} {Structure-assigned optical spectra of single-walled carbon
  nanotubes},\ }\href {https://doi.org/10.1126/science.1078727} {\bibfield
  {journal} {\bibinfo  {journal} {Science}\ }\textbf {\bibinfo {volume}
  {298}},\ \bibinfo {pages} {2361} (\bibinfo {year} {2002})}\BibitemShut
  {NoStop}%
\bibitem [{\citenamefont {Slepyan}\ \emph {et~al.}(1998)\citenamefont
  {Slepyan}, \citenamefont {Maksimenko}, \citenamefont {Lakhtakia},
  \citenamefont {Yevtushenko},\ and\ \citenamefont
  {Gusakov}}]{PhysRevB.57.9485}%
  \BibitemOpen
  \bibfield  {author} {\bibinfo {author} {\bibfnamefont {G.~Y.}\ \bibnamefont
  {Slepyan}}, \bibinfo {author} {\bibfnamefont {S.~A.}\ \bibnamefont
  {Maksimenko}}, \bibinfo {author} {\bibfnamefont {A.}~\bibnamefont
  {Lakhtakia}}, \bibinfo {author} {\bibfnamefont {O.~M.}\ \bibnamefont
  {Yevtushenko}},\ and\ \bibinfo {author} {\bibfnamefont {A.~V.}\ \bibnamefont
  {Gusakov}},\ }\bibfield  {title} {\bibinfo {title} {Electronic and
  electromagnetic properties of nanotubes},\ }\href
  {https://doi.org/10.1103/PhysRevB.57.9485} {\bibfield  {journal} {\bibinfo
  {journal} {Physical Review B}\ }\textbf {\bibinfo {volume} {57}},\ \bibinfo
  {pages} {9485} (\bibinfo {year} {1998})}\BibitemShut {NoStop}%
\bibitem [{\citenamefont {Wang}\ \emph {et~al.}(2009)\citenamefont {Wang},
  \citenamefont {Chu}, \citenamefont {Wang},\ and\ \citenamefont
  {Guo}}]{PhysRevB.80.235430}%
  \BibitemOpen
  \bibfield  {author} {\bibinfo {author} {\bibfnamefont {B.}~\bibnamefont
  {Wang}}, \bibinfo {author} {\bibfnamefont {R.}~\bibnamefont {Chu}}, \bibinfo
  {author} {\bibfnamefont {J.}~\bibnamefont {Wang}},\ and\ \bibinfo {author}
  {\bibfnamefont {H.}~\bibnamefont {Guo}},\ }\bibfield  {title} {\bibinfo
  {title} {First-principles calculation of chiral current and quantum
  self-inductance of carbon nanotubes},\ }\href
  {https://doi.org/10.1103/PhysRevB.80.235430} {\bibfield  {journal} {\bibinfo
  {journal} {Physical Review B}\ }\textbf {\bibinfo {volume} {80}},\ \bibinfo
  {pages} {235430} (\bibinfo {year} {2009})}\BibitemShut {NoStop}%
\bibitem [{\citenamefont {Mintmire}\ and\ \citenamefont
  {White}(1998)}]{Mintmire1998}%
  \BibitemOpen
  \bibfield  {author} {\bibinfo {author} {\bibfnamefont {J.~W.}\ \bibnamefont
  {Mintmire}}\ and\ \bibinfo {author} {\bibfnamefont {C.~T.}\ \bibnamefont
  {White}},\ }\bibfield  {title} {\bibinfo {title} {Universal density of states
  for carbon nanotubes},\ }\href {https://doi.org/10.1103/PhysRevLett.81.2506}
  {\bibfield  {journal} {\bibinfo  {journal} {Physical Review Letters}\
  }\textbf {\bibinfo {volume} {81}},\ \bibinfo {pages} {2506} (\bibinfo {year}
  {1998})}\BibitemShut {NoStop}%
\bibitem [{\citenamefont {Palacios}\ \emph {et~al.}(2003)\citenamefont
  {Palacios}, \citenamefont {P\'erez-Jim\'enez}, \citenamefont {Louis},
  \citenamefont {SanFabi\'an},\ and\ \citenamefont
  {Verg\'es}}]{PhysRevLett.90.106801}%
  \BibitemOpen
  \bibfield  {author} {\bibinfo {author} {\bibfnamefont {J.~J.}\ \bibnamefont
  {Palacios}}, \bibinfo {author} {\bibfnamefont {A.~J.}\ \bibnamefont
  {P\'erez-Jim\'enez}}, \bibinfo {author} {\bibfnamefont {E.}~\bibnamefont
  {Louis}}, \bibinfo {author} {\bibfnamefont {E.}~\bibnamefont {SanFabi\'an}},\
  and\ \bibinfo {author} {\bibfnamefont {J.~A.}\ \bibnamefont {Verg\'es}},\
  }\bibfield  {title} {\bibinfo {title} {First-principles phase-coherent
  transport in metallic nanotubes with realistic contacts},\ }\href
  {https://doi.org/10.1103/PhysRevLett.90.106801} {\bibfield  {journal}
  {\bibinfo  {journal} {Physical Review Letters}\ }\textbf {\bibinfo {volume}
  {90}},\ \bibinfo {pages} {106801} (\bibinfo {year} {2003})}\BibitemShut
  {NoStop}%
\bibitem [{\citenamefont {Nemec}\ \emph {et~al.}(2008)\citenamefont {Nemec},
  \citenamefont {Tom\'anek},\ and\ \citenamefont {Cuniberti}}]{Nemec}%
  \BibitemOpen
  \bibfield  {author} {\bibinfo {author} {\bibfnamefont {N.}~\bibnamefont
  {Nemec}}, \bibinfo {author} {\bibfnamefont {D.}~\bibnamefont {Tom\'anek}},\
  and\ \bibinfo {author} {\bibfnamefont {G.}~\bibnamefont {Cuniberti}},\
  }\bibfield  {title} {\bibinfo {title} {Modeling extended contacts for
  nanotube and graphene devices},\ }\href
  {https://doi.org/10.1103/PhysRevB.77.125420} {\bibfield  {journal} {\bibinfo
  {journal} {Physical Review B}\ }\textbf {\bibinfo {volume} {77}},\ \bibinfo
  {pages} {125420} (\bibinfo {year} {2008})}\BibitemShut {NoStop}%
\bibitem [{\citenamefont {Chen}\ \emph {et~al.}(2026)\citenamefont {Chen},
  \citenamefont {Hod}, \citenamefont {Gersten},\ and\ \citenamefont
  {Nitzan}}]{doi:10.1021/acs.jctc.5c01410}%
  \BibitemOpen
  \bibfield  {author} {\bibinfo {author} {\bibfnamefont {Y.}~\bibnamefont
  {Chen}}, \bibinfo {author} {\bibfnamefont {O.}~\bibnamefont {Hod}}, \bibinfo
  {author} {\bibfnamefont {J.}~\bibnamefont {Gersten}},\ and\ \bibinfo {author}
  {\bibfnamefont {A.}~\bibnamefont {Nitzan}},\ }\bibfield  {title} {\bibinfo
  {title} {Chirality-induced orbital-angular-momentum selectivity in electron
  transmission and scattering},\ }\href
  {https://doi.org/10.1021/acs.jctc.5c01410} {\bibfield  {journal} {\bibinfo
  {journal} {Journal of Chemical Theory and Computation}\ }\textbf {\bibinfo
  {volume} {22}},\ \bibinfo {pages} {20} (\bibinfo {year} {2026})}\BibitemShut
  {NoStop}%
\bibitem [{\citenamefont {Chen}\ and\ \citenamefont
  {Hod}(2023)}]{10.1063/5.0156491}%
  \BibitemOpen
  \bibfield  {author} {\bibinfo {author} {\bibfnamefont {Y.}~\bibnamefont
  {Chen}}\ and\ \bibinfo {author} {\bibfnamefont {O.}~\bibnamefont {Hod}},\
  }\bibfield  {title} {\bibinfo {title} {Chirality induced spin selectivity: A
  classical spin-off},\ }\href {https://doi.org/10.1063/5.0156491} {\bibfield
  {journal} {\bibinfo  {journal} {The Journal of Chemical Physics}\ }\textbf
  {\bibinfo {volume} {158}},\ \bibinfo {pages} {244102} (\bibinfo {year}
  {2023})}\BibitemShut {NoStop}%
\bibitem [{\citenamefont {Trocha}(2010)}]{PhysRevB.82.115320}%
  \BibitemOpen
  \bibfield  {author} {\bibinfo {author} {\bibfnamefont {P.}~\bibnamefont
  {Trocha}},\ }\bibfield  {title} {\bibinfo {title} {Beating in electronic
  transport through quantum dot based devices},\ }\href
  {https://doi.org/10.1103/PhysRevB.82.115320} {\bibfield  {journal} {\bibinfo
  {journal} {Physical Review B}\ }\textbf {\bibinfo {volume} {82}},\ \bibinfo
  {pages} {115320} (\bibinfo {year} {2010})}\BibitemShut {NoStop}%
\bibitem [{\citenamefont {Souza}(2007)}]{PhysRevB.76.205315}%
  \BibitemOpen
  \bibfield  {author} {\bibinfo {author} {\bibfnamefont {F.~M.}\ \bibnamefont
  {Souza}},\ }\bibfield  {title} {\bibinfo {title} {Spin-dependent ringing and
  beats in a quantum dot system},\ }\href
  {https://doi.org/10.1103/PhysRevB.76.205315} {\bibfield  {journal} {\bibinfo
  {journal} {Physical Review B}\ }\textbf {\bibinfo {volume} {76}},\ \bibinfo
  {pages} {205315} (\bibinfo {year} {2007})}\BibitemShut {NoStop}%
\bibitem [{\citenamefont {Barros}\ \emph {et~al.}(2006)\citenamefont {Barros},
  \citenamefont {Jorio}, \citenamefont {Samsonidze}, \citenamefont {Capaz},
  \citenamefont {{Souza Filho}}, \citenamefont {{Mendes Filho}}, \citenamefont
  {Dresselhaus},\ and\ \citenamefont {Dresselhaus}}]{BARROS2006261}%
  \BibitemOpen
  \bibfield  {author} {\bibinfo {author} {\bibfnamefont {E.~B.}\ \bibnamefont
  {Barros}}, \bibinfo {author} {\bibfnamefont {A.}~\bibnamefont {Jorio}},
  \bibinfo {author} {\bibfnamefont {G.~G.}\ \bibnamefont {Samsonidze}},
  \bibinfo {author} {\bibfnamefont {R.~B.}\ \bibnamefont {Capaz}}, \bibinfo
  {author} {\bibfnamefont {A.~G.}\ \bibnamefont {{Souza Filho}}}, \bibinfo
  {author} {\bibfnamefont {J.}~\bibnamefont {{Mendes Filho}}}, \bibinfo
  {author} {\bibfnamefont {G.}~\bibnamefont {Dresselhaus}},\ and\ \bibinfo
  {author} {\bibfnamefont {M.~S.}\ \bibnamefont {Dresselhaus}},\ }\bibfield
  {title} {\bibinfo {title} {Review on the symmetry-related properties of
  carbon nanotubes},\ }\href
  {https://doi.org/https://doi.org/10.1016/j.physrep.2006.05.007} {\bibfield
  {journal} {\bibinfo  {journal} {Physics Reports}\ }\textbf {\bibinfo {volume}
  {431}},\ \bibinfo {pages} {261} (\bibinfo {year} {2006})}\BibitemShut
  {NoStop}%
\bibitem [{\citenamefont {Streib}(2021)}]{PhysRevB.103.L100409}%
  \BibitemOpen
  \bibfield  {author} {\bibinfo {author} {\bibfnamefont {S.}~\bibnamefont
  {Streib}},\ }\bibfield  {title} {\bibinfo {title} {Difference between angular
  momentum and pseudoangular momentum},\ }\href
  {https://doi.org/10.1103/PhysRevB.103.L100409} {\bibfield  {journal}
  {\bibinfo  {journal} {Physical Review B}\ }\textbf {\bibinfo {volume}
  {103}},\ \bibinfo {pages} {L100409} (\bibinfo {year} {2021})}\BibitemShut
  {NoStop}%
\bibitem [{\citenamefont {White}\ \emph {et~al.}(1993)\citenamefont {White},
  \citenamefont {Robertson},\ and\ \citenamefont {Mintmire}}]{Mintmire}%
  \BibitemOpen
  \bibfield  {author} {\bibinfo {author} {\bibfnamefont {C.~T.}\ \bibnamefont
  {White}}, \bibinfo {author} {\bibfnamefont {D.~H.}\ \bibnamefont
  {Robertson}},\ and\ \bibinfo {author} {\bibfnamefont {J.~W.}\ \bibnamefont
  {Mintmire}},\ }\bibfield  {title} {\bibinfo {title} {Helical and rotational
  symmetries of nanoscale graphitic tubules},\ }\href
  {https://doi.org/10.1103/PhysRevB.47.5485} {\bibfield  {journal} {\bibinfo
  {journal} {Physical Review B}\ }\textbf {\bibinfo {volume} {47}},\ \bibinfo
  {pages} {5485(R)} (\bibinfo {year} {1993})}\BibitemShut {NoStop}%
\bibitem [{\citenamefont {Mintmire}\ and\ \citenamefont
  {White}(1995)}]{MINTMIRE1995893}%
  \BibitemOpen
  \bibfield  {author} {\bibinfo {author} {\bibfnamefont {J.}~\bibnamefont
  {Mintmire}}\ and\ \bibinfo {author} {\bibfnamefont {C.}~\bibnamefont
  {White}},\ }\bibfield  {title} {\bibinfo {title} {Electronic and structural
  properties of carbon nanotubes},\ }\href
  {https://doi.org/https://doi.org/10.1016/0008-6223(95)00018-9} {\bibfield
  {journal} {\bibinfo  {journal} {Carbon}\ }\textbf {\bibinfo {volume} {33}},\
  \bibinfo {pages} {893} (\bibinfo {year} {1995})},\ \bibinfo {note}
  {nanotubes}\BibitemShut {NoStop}%
\bibitem [{\citenamefont {Lunde}\ \emph {et~al.}(2005)\citenamefont {Lunde},
  \citenamefont {Flensberg},\ and\ \citenamefont {Jauho}}]{Lunde}%
  \BibitemOpen
  \bibfield  {author} {\bibinfo {author} {\bibfnamefont {A.~M.}\ \bibnamefont
  {Lunde}}, \bibinfo {author} {\bibfnamefont {K.}~\bibnamefont {Flensberg}},\
  and\ \bibinfo {author} {\bibfnamefont {A.-P.}\ \bibnamefont {Jauho}},\
  }\bibfield  {title} {\bibinfo {title} {Intershell resistance in multiwall
  carbon nanotubes: A coulomb drag study},\ }\href
  {https://doi.org/10.1103/PhysRevB.71.125408} {\bibfield  {journal} {\bibinfo
  {journal} {Physical Review B}\ }\textbf {\bibinfo {volume} {71}},\ \bibinfo
  {pages} {125408} (\bibinfo {year} {2005})}\BibitemShut {NoStop}%
\bibitem [{\citenamefont {Allen}\ \emph {et~al.}(1992)\citenamefont {Allen},
  \citenamefont {Beijersbergen}, \citenamefont {Spreeuw},\ and\ \citenamefont
  {Woerdman}}]{Allen1992}%
  \BibitemOpen
  \bibfield  {author} {\bibinfo {author} {\bibfnamefont {L.}~\bibnamefont
  {Allen}}, \bibinfo {author} {\bibfnamefont {M.~W.}\ \bibnamefont
  {Beijersbergen}}, \bibinfo {author} {\bibfnamefont {R.~J.~C.}\ \bibnamefont
  {Spreeuw}},\ and\ \bibinfo {author} {\bibfnamefont {J.~P.}\ \bibnamefont
  {Woerdman}},\ }\bibfield  {title} {\bibinfo {title} {Orbital angular momentum
  of light and the transformation of laguerre-gaussian laser modes},\ }\href
  {https://doi.org/10.1103/PhysRevA.45.8185} {\bibfield  {journal} {\bibinfo
  {journal} {Physical Review A}\ }\textbf {\bibinfo {volume} {45}},\ \bibinfo
  {pages} {8185} (\bibinfo {year} {1992})}\BibitemShut {NoStop}%
\bibitem [{\citenamefont {Slater}\ and\ \citenamefont
  {Koster}(1954)}]{SlaterKoster}%
  \BibitemOpen
  \bibfield  {author} {\bibinfo {author} {\bibfnamefont {J.~C.}\ \bibnamefont
  {Slater}}\ and\ \bibinfo {author} {\bibfnamefont {G.~F.}\ \bibnamefont
  {Koster}},\ }\bibfield  {title} {\bibinfo {title} {Simplified {LCAO} method
  for the periodic potential problem},\ }\href
  {https://doi.org/10.1103/PhysRev.94.1498} {\bibfield  {journal} {\bibinfo
  {journal} {Physical Review}\ }\textbf {\bibinfo {volume} {94}},\ \bibinfo
  {pages} {1498} (\bibinfo {year} {1954})}\BibitemShut {NoStop}%
\bibitem [{\citenamefont {Kleiner}\ and\ \citenamefont
  {Eggert}(2001)}]{KleinerEggert2001}%
  \BibitemOpen
  \bibfield  {author} {\bibinfo {author} {\bibfnamefont {A.}~\bibnamefont
  {Kleiner}}\ and\ \bibinfo {author} {\bibfnamefont {S.}~\bibnamefont
  {Eggert}},\ }\bibfield  {title} {\bibinfo {title} {Band gaps of primary
  metallic carbon nanotubes},\ }\href
  {https://doi.org/10.1103/PhysRevB.63.073408} {\bibfield  {journal} {\bibinfo
  {journal} {Physical Review B}\ }\textbf {\bibinfo {volume} {63}},\ \bibinfo
  {pages} {073408} (\bibinfo {year} {2001})}\BibitemShut {NoStop}%
\end{thebibliography}%

\end{document}